\documentclass[useAMS,usenatbib]{mn2e}

 \usepackage{Times}

\usepackage{psfrag}
\usepackage{pspicture}
\usepackage{graphicx}

\def\gsim{~\lower.6ex\hbox{$\buildrel < \over \sim$}~}

\title[Star formation, environment \& stellar mass at $\bf z\sim1$]{The dependence of star formation activity on environment and stellar mass at $\bf z\sim1$ from the HiZELS-H$\alpha$ survey \thanks{Based on observations obtained using the Wide Field CAMera on the 3.8m United Kingdom Infrared Telescope, as part of the High-$z$ Emission Line Survey (HiZELS).} }
\author[D. Sobral et al.]{David Sobral$^{1}$\thanks{E-mail: drss@roe.ac.uk}, Philip N. Best$^{1}$, Ian Smail$^{2}$, James E. Geach$^{2}$, Michele Cirasuolo$^{1,3}$,
\newauthor Timothy Garn$^{1}$ and Gavin B. Dalton$^{4,5}$\\
$^{1}$SUPA, Institute for Astronomy, Royal Observatory of Edinburgh, Blackford Hill, Edinburgh, EH9 3HJ, UK\\
$^{2}$Institute for Computational Cosmology, Durham University, South Road, Durham, DH1 3LE, UK\\
$^{3}$Astronomy Technology Centre, Royal Observatory of Edinburgh, Blackford Hill, Edinburgh, EH9 3HJ, UK\\
$^{4}$Astrophysics, Department of Physics, Keble Road, Oxford, OX1 3RH, UK\\
$^{5}$Space Science and Technology, Rutherford Appleton Laboratory, HSIC, Didcot, OX11 0QX, UK}
\begin{document}

\date{Accepted 2010 September 14. Received August 2010 25; in original form 2010 July 15}

\pagerange{\pageref{firstpage}--\pageref{lastpage}} \pubyear{2010}

\maketitle

\label{firstpage}

\begin{abstract}

This paper presents an environment and stellar mass study of a large sample of star-forming H$\alpha$ emitters at $z = 0.84$ from the High-$z$ Emission Line Survey (HiZELS), over 1.3 deg$^2$ split over two fields (COSMOS and UKIDSS UDS). By taking advantage of a truly panoramic coverage of a wide range of environments, from the field to a rich cluster, it is shown that both stellar mass and environment play crucial roles in determining the properties of star-forming galaxies. Specific star formation rates (sSFRs) decline with stellar mass in all environments, and the fraction of H$\alpha$ star-forming galaxies declines sharply from $\approx40$ per cent for galaxies with masses around $10^{10}$\,M$_{\odot}$ to effectively zero above $10^{11.5}$\,M$_{\odot}$, confirming that mass-downsizing is generally in place by $z\sim1$. The fraction of star-forming galaxies is also found to fall sharply as a function of local environmental density from $\approx40$ per cent in the field to approaching zero at rich group/cluster densities. When star formation does occur in such high density regions, it is found to be mostly dominated by potential mergers and, indeed, if only non-merging star-forming galaxies are considered, then the environment and mass trends are even stronger and are qualitatively similar at all masses and environments, respectively, as in the local Universe. The median star-formation rate of H$\alpha$ emitters at $z=0.84$ is found to increase with density for both field and intermediate (group or cluster outskirts) densities; this is clearly seen as a change in the faint-end slope of the H$\alpha$ luminosity function from steep ($\alpha \approx -1.9$), in poor fields, to shallow ($\alpha \approx -1.1$) in groups and clusters. Interestingly, the relation between median SFR and environment is only found for low to moderate-mass galaxies (with stellar masses below about $10^{10.6}$\,M$_{\odot}$), and is not seen for the most massive star-forming galaxies. Overall, these observations provide a detailed view over a sufficiently large range of mass and environment to reconcile previous observational claims: stellar mass is the primary predictor of star-formation activity at $z\sim1$, but the environment, while initially enhancing the median SFR of (lower-mass) star-forming galaxies, is ultimately responsible for suppressing star-formation activity in all galaxies above surface densities of 10-30 Mpc$^{-2}$ (group and cluster environments).

\end{abstract}


\section{Introduction}\label{intro}

At the present epoch, star formation activity is strongly dependent on environment \citep[e.g.][]{lewiss,Gomez,Tanaka04,Mahajan2010}: while clusters of galaxies seem to be primarily populated by passively-evolving galaxies, star-forming galaxies are mainly found in low-density environments \citep{Dressler80}. It is well-established \citep[e.g.][]{Gomez,Kauff,Best2004} that the typical star formation rates of galaxies -- and the star-forming fraction -- decrease with local galaxy density (often projected local density, $\Sigma$) both in the local Universe and at moderate redshift \citep[e.g.][]{Kodama04}. Indeed, at $z\sim0.2-0.3$, studies such as \cite{Couch01} or \cite{Balogh02} found that the H$\alpha$ luminosity function in rich clusters, whilst having roughly the same shape as in the field, have a much lower normalisation ($\sim50$ per cent), consistent with a significant suppression of star formation in such environments.

Typically, active star-forming galaxies in the local Universe are also found to have lower masses than passive galaxies and, indeed, the most massive galaxies are mostly non-star-forming \citep[an observational result often known as mass-downsizing;][]{Cowie}; this implies that the most massive galaxies assembled their stellar mass quicker and have not formed a significant amount of stars since then. Massive galaxies are predominantly found in high density environments, but the mass-downsizing trend is not simply a consequence of the environmental dependence, nor vice-versa. Recent studies such as \cite{Mahajan2010} -- and references therein -- show that the environment trends hold for both massive and dwarf galaxies, and that the mass trends also hold for both field and cluster environments. \cite{Peng} also finds that the effects of mass and environment are completely separable.

When did the environment and mass dependences of star-forming galaxies start to be observable, and how did they affect the evolution of galaxies and clusters? Clearly, in order to properly answer such questions it is mandatory to conduct observational surveys at high redshift, which can then be used to test theoretical models of galaxy evolution. By $z\sim1$, some authors have claimed to have found a flattening or even a definitive reverse of the relation between star formation activity and local galaxy density, either by studying how the average/median star formation rates of galaxies change with local density \citep[e.g.][]{elbaz,cooper} or by looking at the star-forming fraction as a function of density \citep[e.g.][]{Ideue}. Similar trends have also been suggested at $z\sim2$ by \cite{Kodama}, and recently \cite{Hayashi09} found that the fraction of star-forming galaxies seems to remain constant up to the highest densities in a cluster environment at $z\sim1.5$. These results would be naturally interpreted as a sign of a clear evolution if other studies \citep[e.g.][]{Finn05,Patel,Koyama09} had not indicated the opposite result. The latter studies argue that even at $z\sim1$, and for the richest/cluster environments, both star formation rate and the star-forming fraction decline with increasing local density. Nevertheless, \cite{Koyama09} also found that the H$\alpha$ luminosity function of a rich cluster at $z\sim0.8$ has roughly the same shape as the field luminosity function from \cite{Sobral09}, but with a higher normalisation; this implies that even if the environment is already suppressing star formation by $z\sim1$, there is still a considerable population of star-forming galaxies in high-redshift clusters \citep[c.f.][]{ButcherOemler}.

Part of the discrepancies in environmental trends between different studies may well be connected with possible mass dependences already in place at high redshift. While at $z\sim0$ the large and deep surveys can now probe very different mass regimes with large enough samples within each environment to confirm and distinguish mass and environment trends individually \citep[c.f.][]{Peng}, that is clearly not the case at high redshift, where luminosity limits inevitably bias samples towards higher mass galaxies. Thus, if mass-downsizing is already in place at high redshift, and massive galaxies are preferentially found in high density regions, samples with different luminosity limits and/or based on different selections may find contradictory results. Indeed, such downsizing trends have now been identified out to $z\sim1.5$, by revealing that the most massive galaxies are undergoing very little or no star formation, or by finding a population of massive galaxies already in the red sequence \citep[e.g.][]{Bauer05,Marchesini09,santini2009,Fontanot,Cirasuolo10}. Therefore, in order to identify and distinguish the separate roles of mass and environment on star formation at high redshift, one really requires clean, robust and large samples of star-forming galaxies residing in a wide range of environments and with a wide range of masses, together with samples of the underlying population found at the same redshift.

Narrow-band H$\alpha$ surveys are one of the most effective ways to gather representative samples of star-forming galaxies at different epochs, and the scientific potential of these is now being widely explored, following the development of wide-field cameras in the near-infrared. This has led to significant progress \citep[e.g.][]{Ly2007,Shioya2008,G08,Sobral09,Garn09,Sobral09c} taking advantage of large samples of typically hundreds of H$\alpha$ emitters up to $z=2.23$. In particular, HiZELS, the Hi-Redshift($z$) Emission Line Survey \citep[][]{G08,Sobral09}, is playing a fundamental role by obtaining and exploring unique samples of star-forming galaxies selected through their H$\alpha$ emission \citep[and other major emission lines;][]{Sobral09b}, redshifted into the $J$, $H$ and $K$ bands \citep[see][for an overview]{BEST10}. By using a set of existing and custom-made narrow-band filters, HiZELS is surveying H$\alpha$ emitters at $z=0.84$, $z=1.47$ and $z=2.23$ over several square degree areas of extragalactic sky. Narrow-band surveys such as HiZELS probe remarkably thin redshift slices ($\Delta z\approx0.03$), allowing the study of very well-defined cosmic epochs. They survey the redshift slices down to a known flux limit and they can select equivalent populations at different redshifts. Combined with a wealth of  high-quality multi-wavelength data in the targeted fields, wide narrow-band surveys like HiZELS can be fully explored to detail the roles of ``nature'' and ``nurture'' in galaxy evolution, by unveiling the importance of the environment on a well-defined star-forming population at a single epoch. 

This paper presents a detailed study of the dependencies of star formation on environment and mass, and their relations with galaxy morphology, using a large sample of narrow-band selected H$\alpha$ emitters at $z=0.84$. Using this sample, it is possible to reconcile many of the apparently contradictory results in the literature. The paper is organised in the following way. \S2 outlines the data and selection, the samples, their sky distribution and the mass and local density estimations. \S3 presents the dependence of star formation activity on mass, together with the downsizing trends found at $z=0.84$. \S4 details the connection between star formation and the environment, presenting the relations between local density, star formation activity and star-forming fraction for H$\alpha$ emitters at $z=0.84$. \S5 presents a detailed, combined mass-environment view of star formation at $z\sim1$. \S6 discusses and interprets the results in the context of other studies and finally \S7 gives a summary and the conclusions. An H$_0=70$\,km\,s$^{-1}$\,Mpc$^{-1}$, $\Omega_M=0.3$ and $\Omega_{\Lambda}=0.7$ cosmology is used and, except where otherwise noted, magnitudes are presented in the AB system.

\section{SAMPLES AND PROPERTIES}\label{data_technique}

\subsection{The sample of robust H$\alpha$ emitters at $\bf z=0.84$}\label{sample}

This study uses the large sample of H$\alpha$ emitters at $z=0.84$ from HiZELS presented in Sobral et al. (2009a, S09a). Briefly, the sample was derived from a narrow-band $J$ filter ($\lambda_{\rm eff} = 1.211\umu$m, $\Delta\lambda=0.015\umu$m) survey using the Wide Field CAMera (WFCAM) on the United Kingdom Infrared Telescope (UKIRT), reaching a flux limit of $8\times10^{-17}$\,erg\,s$^{-1}$\,cm$^{-2}$ over $\sim 0.7\deg^2$ in the UKIDSS Ultra Deep Survey \citep[UDS,][]{2007MNRAS.379.1599L} field, and $\sim 0.8\deg^2$ in the Cosmological Evolution Survey \citep[COSMOS,][]{2007ApJS..172....1S} field. The survey resulted in the selection of 1517 line emitters which were clearly detected in NB$_{\rm J}$ (signal-to-noise ratio $>3$) with a $J$-NB$_{\rm J}$ colour excess significance of $\Sigma>2.5$ and observed equivalent width EW$ >50$ \AA \ -- see S09a for full details on the selection.

Photometric redshifts\footnote{Photometric redshifts used in S09a for COSMOS present $\sigma(\Delta z ) = 0.03$, where $\Delta z$ = $(z_{phot}-z_{spec})/(1+z_{spec})$; the fraction of outliers, defined as sources with $\Delta  z>3\sigma(\Delta z)$, is lower than 3 per cent, while for UDS, the photometric redshifts have $\sigma(\Delta z) = 0.04$, with 2 per cent of outliers.} were used to select a sample of 743 H$\alpha$ emitters over a co-moving volume of $1.8 \times 10^5$\,Mpc$^3$ at $z=0.845$. Of these, 477 are in COSMOS (0.76 deg$^2$) and 266 in UDS (0.54 deg$^2$). The completeness and reliability of this sample were studied using the $\sim10^4$ available redshifts from $z$COSMOS Data Release 2 \citep{zCOSMOS}, spectroscopically confirming $\sim100$ H$\alpha$ emitters within the photometric redshift selected sample; this allowed to estimate a $>95$ per cent reliability and $>96$ per cent completeness for the sample in COSMOS. Here, the sample of H$\alpha$ emitters is modified from that in S09a in three ways:

\noindent (i) Following \cite{Sobral09c}, the H$\alpha$ sample is slightly modified on the basis of the new photometric redshifts, incorporating a higher number of bands, deeper data and accounting for possible emission-line flux contamination of the broad-bands \citep[c.f.][Cirasuolo et al. in prep.]{Ilbert09}. In particular, the revised sample in UDS contains 257 H$\alpha$ emitters over 0.52 deg$^2$, whilst that in COSMOS is unchanged (477 emitters over 0.76 deg$^2$).

\noindent(ii) Some of the H$\alpha$ emitters may contain an AGN, which can be responsible for a significant fraction of the H$\alpha$ flux and are capable of changing the host galaxy spectral energy distribution. Within the H$\alpha$ sample, AGN candidates have been identified by \cite{Garn09} using a wide range of methods. The analysis led to the identification of a maximum of 74 AGN (40 in COSMOS and 34 in UDS). From these, 40 are classified as possible AGN and the remaining 34 as likely AGN -- implying $5-11$ per cent AGN contamination of the H$\alpha$ sample at $z=0.84$. For the analysis presented in this paper all 74 AGN candidates (possible and likely) are excluded from the star-forming H$\alpha$ sample (though they are kept in the underlying population sample), but it has been checked that if these are retained then none of the results or conclusions of this paper are altered. This results in a robust sample of 660 star-forming H$\alpha$ emitters at $z=0.84$.

\noindent(iii) Finally, as detailed in \S2.3, the sample is also modified to include potential H$\alpha$ emitters with low equivalent widths.

\subsection{The underlying population samples} \label{und_pop}

As the samples were obtained in two of the best-studied square degree areas of equatorial sky, a wealth of multi-wavelength data are available. These include deep intermediate and broad-band imaging from the ultra-violet (UV) to the infrared (IR), making it possible for excellent photometric redshifts (photo-$z$s) to be determined. These data are fundamental to test downsizing and environmental trends, since these trends can only be investigated by selecting an underlying population of galaxies within the same redshift range as the H$\alpha$ population. Whilst the underlying population will never be as clean as the H$\alpha$ sample, the high quality photometric redshifts at $z\sim0.8$ available in COSMOS and UDS are good enough to do a first-pass cut at emulating the narrow-band filter selection. For the COSMOS field, the $z$COSMOS DR2 secure ($>99$ per cent confidence) spectroscopic redshifts can then be used to study the completeness and the contamination of various photometric-redshift selected samples (as well as to improve both, by taking out confirmed contaminants and introducing spectroscopically confirmed galaxies missed by the photometric redshifts). These results provide appropriate corrections to the photometric redshift selection. In Appendix A, a range of different initial photo-$z$ cuts are considered and the resultant correction factors based on the contamination and completeness are calculated; these are used to show that the results in the paper are robust to changes in the photometric redshift cut.

The analysis in this paper is carried out using the background population sample with $0.82<z_{photo}<0.87$ for COSMOS (where photo-$z$s are very precise; the sample is also corrected and improved using spectroscopic redshifts), estimated to be 77 per cent complete and with a contamination of 59 per cent (when compared with the redshift range covered by the narrow-band top hat filter profile, $0.83<z<0.86$). For UDS, the underlying sample is obtained with galaxies having $0.79<z_{photo}<0.9$. Reliable contamination and completeness estimates are not available due to the lack of large spectroscopic data-sets, but the completeness can still be roughly estimated based on the recovery of H$\alpha$ emitters (see Appendix A), yielding $\sim60$\,per cent. Also, given the uncertainty on the photo-$z$s, a $\sim75$ per cent contamination is expected and this is assumed throughout the analysis. This, of course, results in a higher uncertainty in the following analysis for UDS, but as the next sections will fully detail, similar trends are recovered both in COSMOS and UDS.

A common $K<23$ (observed, corresponding to rest-frame $J$ at $z=0.84$) cut is applied (prior to the completeness and contamination studies) to match the $K$ completeness of the H$\alpha$ emitters sample and minimize potential biases towards lower luminosity, lower stellar mass galaxies; this also guarantees that both the Cirasuolo et al. (in prep.) and the \cite{Ilbert09} catalogues are more than 90 per cent complete and that the photometric redshifts perform to their best at the studied redshift, avoiding introducing extra biases into the analysis. Sources clearly classified as stars are also rejected. The final underlying sample in COSMOS contains 3656 sources ($\approx4800$ per square degree) including all H$\alpha$ emitters, while the equivalent sample in UDS contains 2688 galaxies ($\approx5100$ per square degree).

\subsection{The potential H$\alpha$ emitters at $\bf z=0.84$ with EW$<50$\,\AA} \label{EW_low_50}

The equivalent width (EW) limit was adopted by S09a to improve the selection of emission-line objects and, in particular, H$\alpha$ emitters at $z=0.84$. It helps to avoid contamination both from bright sources, which can have an apparently significant narrow-band excess due to small colour terms, and from weak (low EW) emission lines, which may be close in wavelength to H$\alpha$ and thus difficult to distinguish with photo-$z$s. However, the EW limit can lead to neglecting galaxies which have emission-line fluxes above the limit of the survey but have a relatively luminous stellar continuum in the $J$ band (rest-frame $R$ at $z=0.84$). In S09a, this is dealt with by conducting detailed simulations which replicate the selection criteria. Simulations are used to compute a completeness correction as a function of luminosity for deriving the H$\alpha$ luminosity function (S09a). These estimate that a total of $\approx10$ per cent of all H$\alpha$ emitters above the flux limit may be missed by the EW$>50$\,\AA \ selection in both COSMOS and UDS. However, they do not provide information about how this varies as a function of mass or environment.

Motivated by the need to provide a fair comparison of the H$\alpha$ selected sample with the underlying population at the same redshift, the latter samples are used (in COSMOS and UDS). For all photo-$z$ selected galaxies, data from S09a are used and the EW\,$>50$ \AA \ requirement is waived. Galaxies within the photo-$z$ sample that present emission line fluxes (colour excesses $\Sigma>2.5$, following S09a) above the survey limit are considered to be potential additional low EW H$\alpha$ emitters.

This approach results in selecting 76 additional sources in COSMOS and 42 in UDS above a flux limit of $8\times10^{-17}$\,erg\,s$^{-1}$\,cm$^{-2}$ (these sources have EW\,$=35\pm11$\,\AA, with the lowest EW being 13\,\AA). For COSMOS, 25 of the 76 recovered sources have high quality spectroscopic redshifts and 17 are indeed confirmed to be H$\alpha$ emitters at $z=0.84$\footnote{The data also reveal 1 [Ar{\sc iii}] emitter at $z=0.69$, 2 [S{\sc ii}] emitters at $z=0.79$, 1 [N{\sc ii}] emitter ($z=0.82$), and 3 [O{\sc i}] emitters at $z=0.89$, while 1 source is not identified ($z\approx0.99$).}, implying a contamination of 32 per cent (clearly demonstrating the initial motivation for the EW limit). After removing the confirmed contaminants, the sample is reduced to 68 H$\alpha$ emitters. Furthermore, inspection of $z$COSMOS spectra of the confirmed H$\alpha$ emitters reveals [O{\sc ii}] emission lines with fluxes $\approx1\times10^{-17}$\,erg\,s$^{-1}$\,cm$^{-2}$, consistent with the estimated weak H$\alpha$ emission for these sources. None of the sources present H$\beta$ or [O{\sc iii}] with $S/N>1$, although 2 of these sources present a red, detectable continuum, with strong absorption lines such as H and K. Only one source is detected as a strong 24 $\umu$m source; this is flagged as an AGN in the general COSMOS catalogue.

For the UDS field, the lack of spectroscopic data does not allow for a similar investigation to be conducted, and with a wider photo-$z$ cut being used, a 40 per cent contamination (by both lines close enough to H$\alpha$ and others, as found in COSMOS) is assumed for the low EW H$\alpha$ sample in UDS.

A sample containing a total of 770 H$\alpha$ emitters will be used in this paper, but the estimated contamination rates of the samples of potential H$\alpha$ emitters are taken into account by using those as relative weights in the analysis. H$\alpha$ luminosities and SFRs for all H$\alpha$ emitters are derived following S09a, accounting for dust extinction as in Garn et al. (2010, equation 10), but it should be noted that the results remain unchanged if no, or a uniform extinction is applied to the sample.

\subsection{Morphological classification with {\it HST} imaging} \label{Morph_class}

The COSMOS field has sensitive {\it Hubble Space Telescope} ({\it HST}) ACS F814W coverage, providing detailed morphological information for all samples. Automated classifications obtained by {\sc zest} (Scarlata et al.\ 2007) are publicly available for the entire COSMOS field, which classifies galaxies in 3 morphological classes: 1) Early-types, 2) Disks/spirals and 3) Irregulars. Reliable estimates exist for 88 per cent of the complete underlying sample in COSMOS at $0.82<z_p<0.87$. Visual classifications were also obtained by S09a for H$\alpha$ emitters at $z=0.84$; the authors find that the two agree very well, with only small differences arising from the use of colour information in S09a (from Subaru data), while {\sc zest} uses the {\it HST} ${F814W}$ imaging only.

Due to a considerable number of sources showing evidence of potential merging activity in S09a, the H$\alpha$ sample was also classified independently into merger classes. Data from Subaru ($B$, $r'$ and $z'$) was combined with the resolution from {\it HST} so that the colour information could be used to obtain a good identification of potential mergers (e.g. distinguishing between normal disks with strong dust bands and real mergers). A source was classified as a potential merger when: 1) it presented a clearly disturbed morphology or disturbed disk, inconsistent with being that of a normal disk galaxy; or 2) it presented more than one bright point-like source and the colour information was inconsistent with one of those being a spiral arm; or 3) there were two or more galaxies which were very close ($<15$\,kpc) -- see S09a for examples of each morphological class. Here, the same criteria are used to extend this classification to galaxies in the sample of potential H$\alpha$ emitters with EW$<50$\,\AA. These sources are found to be mostly discs (65 per cent), with 15 per cent irregulars and 20 per cent early-type galaxies (the significant increase in the fraction of early-type galaxies is the most noticeable difference between the populations with different EWs, as the sample with lower EWs is able to probe much lower specific star formation rates); 33 per cent of the sample shows evidence of potential merging activity.

\begin{table}
 \centering
  \caption{Parameters used to generate the SED templates using the Bruzual \& Charlot (2003) package and the new Bruzual (2007) models.}
  \begin{tabular}{@{}ccc@{}}
  \hline
   \bf $\tau$ (Gyr) & \bf E(B-V) & \bf $Z$ \\
  \hline

0.1& 0 & 0.004  \\
0.3 & 0.1& 0.008 \\
0.5 & 0.2 & 0.02 (Z$_{\odot}$) \\
1.0 & 0.3 &  \\
3.0 & 0.4 & \\
5.0 & 0.5 &  \\
10.0 &  &  \\
  \hline
\end{tabular}
\label{SED_fitting}
\end{table}

\subsection{Stellar mass estimates} \label{MK_mass}

Deep multi-wavelength data are available for all H$\alpha$ emitters and galaxies within the underlying populations and these data are used to compute stellar masses. The multi-wavelength data are used to perform a full SED $\chi^2$ fit -- normalised to one solar mass -- to each galaxy; the stellar-mass is the factor needed to re-scale the luminosities in all bands from the best model to match the observed data. This method is similar to that used by \cite{Fontana04} and \cite{Ilbert2010}. The SED templates are generated with the stellar population synthesis package developed by \cite{BC03}, but the models are drawn from \cite{BC07}. SEDs are produced assuming a universal initial mass function (IMF) from \cite{Chabrier03} and an exponentially declining star formation history with the form $e^{-t/\tau}$, with $\tau$ in the range 0.1 Gyrs to 10 Gyrs. The SEDs were generated for a logarithmic grid of 200 ages (from 0.1 Myr to 6.5 Gyr -- the maximum age at $z=0.84$). Dust extinction was applied to the templates using the \cite{Calzetti00} law with $E(B-V)$ in the range 0 to 0.5 (in steps of 0.1). The models are generated with three different metallicities, including solar; the reader is referred to Table \ref{SED_fitting} for a list of the parameters used.

For the SED fitting, it is assumed that all H$\alpha$ emitters are at $z=0.845$ (the mean redshift for detecting H$\alpha$ with the narrow-band filter)\footnote{The uncertainties in the luminosity distance (and the location of main spectral features) due to the filter profile are random and lead to much smaller uncertainties ($\sim2$ per cent) than other errors and degeneracies in the models.} and the full filter profiles are convolved with the generated SEDs for a direct comparison with the observed total fluxes. Up to 36 wide, medium and narrow bands \citep[from the {\it GALEX} far ultra-violet and near ultra-violet bands to {\it Spitzer} 4 IRAC bands, c.f.][]{Ilbert09,Ilbert2010} are used for the COSMOS field, after applying all of the necessary corrections to obtain the total fluxes, as suggested and fully detailed in Capak et al. (2007) and \cite{Ilbert09,Ilbert2010}: individual aperture corrections, average systematic offsets per band and individual galactic extinction per band. For the UDS field, fewer bands are available (medium band imaging is not available), up to a total of 16, from CFHT $U$ band in the near-ultra-violet to the 4 IRAC bands, but with the advantage of very deep and matched $J$, $H$ and $K$ data from UKIRT/UKIDSS DR5 (c.f. Cirasuolo et al. in prep.). The appropriate corrections (equivalent to those used for COSMOS) are applied to obtain total fluxes in each band for the UDS field. Once all corrections are applied, the COSMOS and UDS magnitude/flux distributions are very well matched.

Stellar masses are similarly derived for all galaxies within the underlying population at $z=0.84$, assuming that they are all at $z=0.845$ for simplicity (and that they are also representative of galaxies at $z\sim0.8$); this thus provides a direct comparison with the masses estimated for the H$\alpha$ population (after taking into account the completeness and contamination of the underlying sample -- see Appendix A).

Finally, the reader should note that even with a large number of photometric data points available, the stellar mass estimate of each individual source is found to be affected by a 1$\sigma$ error (from the multi-dimension $\chi^2$ distribution) of $\sim0.15$ dex, which results from degeneracies between the star formation time-scale $\tau$, age, extinction and, to a smaller extent, metallicity. However, these are the typical errors found in other literature estimates at $z\sim1$, and tests show that the results of this paper are robust against the uncertainties in the stellar mass estimates. 

Other stellar mass estimates in COSMOS have been presented by \cite{Mobash07} \footnote{Improved mass estimates have been obtained by \cite{Ilbert2010} using the similar spectral energy distribution (SED) fitting method used here, but these are not publicly available.}. However, these have been obtained using a much less sophisticated method\footnote{These were obtained from rest-frame $V$ luminosities, $M_V$, and their expected relation with stellar mass using colours drawn from the best spectral-type fit \cite[c.f.][]{Mobash07}.} and are insufficiently accurate to undertake the analysis presented in this paper. Although the mass estimates of Cirasuolo et al. in UDS have been obtained using the same fitting method, there are some differences in the adopted photometric corrections and the SED templates used, so they are re-calculated here to ensure complete consistency between UDS and COSMOS. The reader is referred to Appendix B for a direct comparison between the stellar masses obtained in this paper and those from \cite{Mobash07} and Cirasuolo et al. (in prep.).

\begin{figure}
\centering
\includegraphics[width=8.2cm]{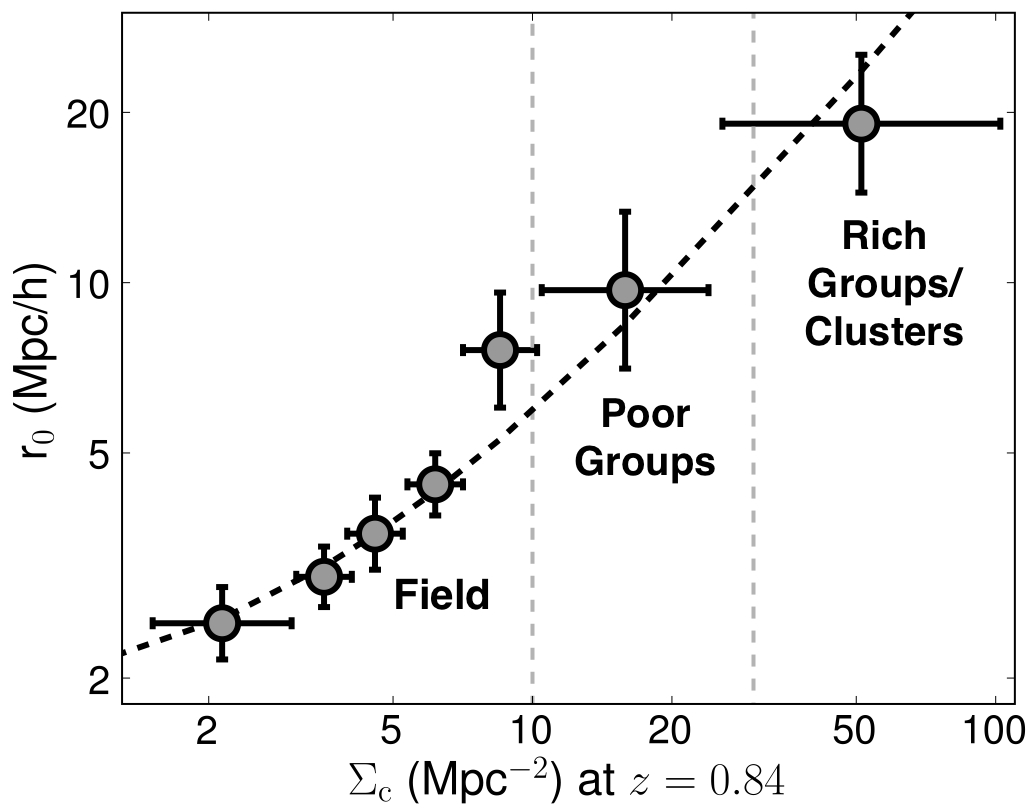}
\caption[wteta_subsam]{The empirical relation between the correlation length, $r_0$, and local density (corrected local projected density), derived for the H$\alpha$ emitters. This provides an intuitive interpretation of both local densities and $r_0$, giving a better insight into the nature of the regions being probed and studied by the survey. This is found to be well fitted by a simple linear fit, given in Section 2.7. Bin sizes are chosen to guarantee that sub-samples have the same number of galaxies.  \label{dens_r0}}
\end{figure}

\subsection{Local density estimates} \label{10th}

Local densities for each source (H$\alpha$ emitters and galaxies in the underlying population samples) are estimated by conducting a 10th nearest neighbour analysis. Briefly, for each source, the angular distances to all other nearby sources are computed. These are then used to determine the projected distance to the 10th-nearest neighbour (in Mpc, using a scale of 7.64 kpc/arcsec at $z=0.845$) and hence to estimate the local projected surface density. Sources near the edges of the fields and/or near masked regions will naturally be biased towards lower local density estimates ($\sim$5 per cent of them are in this situation). Where possible, this is fully tackled by using high-quality photo-$z$ catalogues which exist well beyond the HiZELS NB$_J$ coverage. For regions where that is not possible ($\sim4$ per cent of the UDS emitters are found near the edges of the photo-$z$ data-set or large masked regions, and $<1$ per cent of the COSMOS galaxies lie near masked regions), a simple correction is applied based on the fraction of area lost to the 10th nearest neighbour. Tests show that this does not introduce any bias. 

\begin{figure*}
\centering
\includegraphics[width=17.8cm]{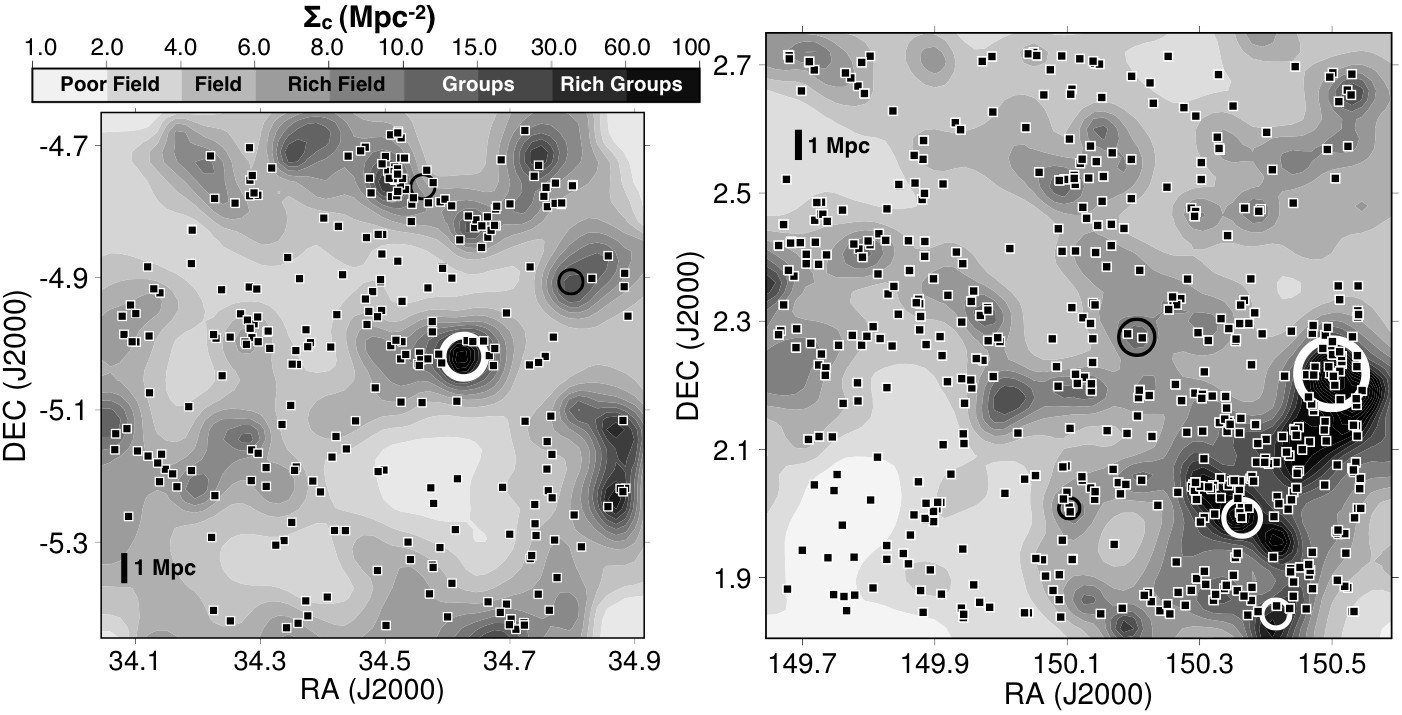}
\caption[on-sky COSMOS_UDS]{The on-sky distribution of the star-forming H$\alpha$ emitters found at $z=0.84$ in the UDS (left panel) and COSMOS (right panel) fields when compared with the local projected density field  which can be linked with qualitative environments -- c.f. Section \ref{dens_clustering}. Sources classed as likely and potential AGN in Garn et al. (2010) are not shown. Circles mark the position of extended X-ray emission from confirmed groups and clusters \citep[from][]{Finoguenov07,FinoguenovUDS} within the narrow-band redshift range, and are scaled to reflect the measured X-ray luminosity (in a log scale). Some lower X-ray luminosity groups are identified with black circles, while richer X-ray groups are plotted in white for good contrast. Note the very rich structure in the COSMOS field (see Section 2.8), providing a unique opportunity to probe the densest environments. \label{on_sky}}
\end{figure*}

Local density estimates derived using the underlying (photo-$z$) sample are corrected by assuming the contamination and completeness fractions derived in Appendix A. In general, given a contamination fraction, Co, a completeness fraction C, and a measured projected surface density $\Sigma$, the corrected local surface density is estimated as:
\begin{equation}
   \Sigma_{\rm c}=\Sigma\times\frac{1-{\rm Co}}{{\rm C}}.
\end{equation} \label{sigma_corr}
When Equation 1 is used to correct densities estimated using different background populations (selected on the basis of photometric redshifts -- see Section \ref{und_pop}) the different estimations are found to agree very well, albeit that the scatter increases for wider (in photo-$z$ cut) underlying population samples (as contamination becomes increasingly dominant. However, the scatter between density measurements obtained using different photo-$z$ samples, or derived using different density estimators (e.g. 5th versus 10th nearest neighbour), is always significantly smaller ($\gsim0.18$ dex) than the bin sizes used in this work ($\approx0.30$ dex). Thus, as Appendix A demonstrates, the main results of the paper do not depend on the choice of the underlying population; the samples used for the analysis are the ones presented in Section \ref{und_pop}.

\subsection{The clustering-density relation and the physical interpretation}\label{dens_clustering}

Estimating local surface densities can quantitatively distinguish higher density regions from lower density regions: but how do the derived surface densities relate to qualitative environments such as fields, groups or even clusters? Here, an attempt to associate qualitative environments with ranges of $\Sigma_{\rm c}$ will be presented, by using the real space correlation length, $r_0$.

Following the procedures fully detailed in Sobral et al. (2010a), the entire sample of H$\alpha$ emitters\footnote{The sample of H$\alpha$ emitters is used instead of the photometric-redshift-selected population as the real-space correlation length can be well estimated (as detailed in Sobral et al. 2010a) without major errors being introduced by the uncertainty in the exact redshift distribution and contamination.} is split based on corrected local surface density, $\Sigma_{\rm c}$, in order to compute the correlation length\footnote{The analysis was carried out fixing $\gamma=-1.8$, since individual sub-samples are too small to accurately determine it. This value of $\gamma$ provides a good fit to the overall sample (though note that, as discussed in Sobral et al. (2010), for narrow-band surveys this does not necessary translate to $\beta=-0.8$ as the Limber approximation would imply, and that the exact transformation between the angular and the 3-D correlation function is used).} of each sub-sample (using the exact relation between the angular and spatial correlation functions; Sobral et al. 2010a).

\begin{figure*}
\centering
\includegraphics[width=16.8cm]{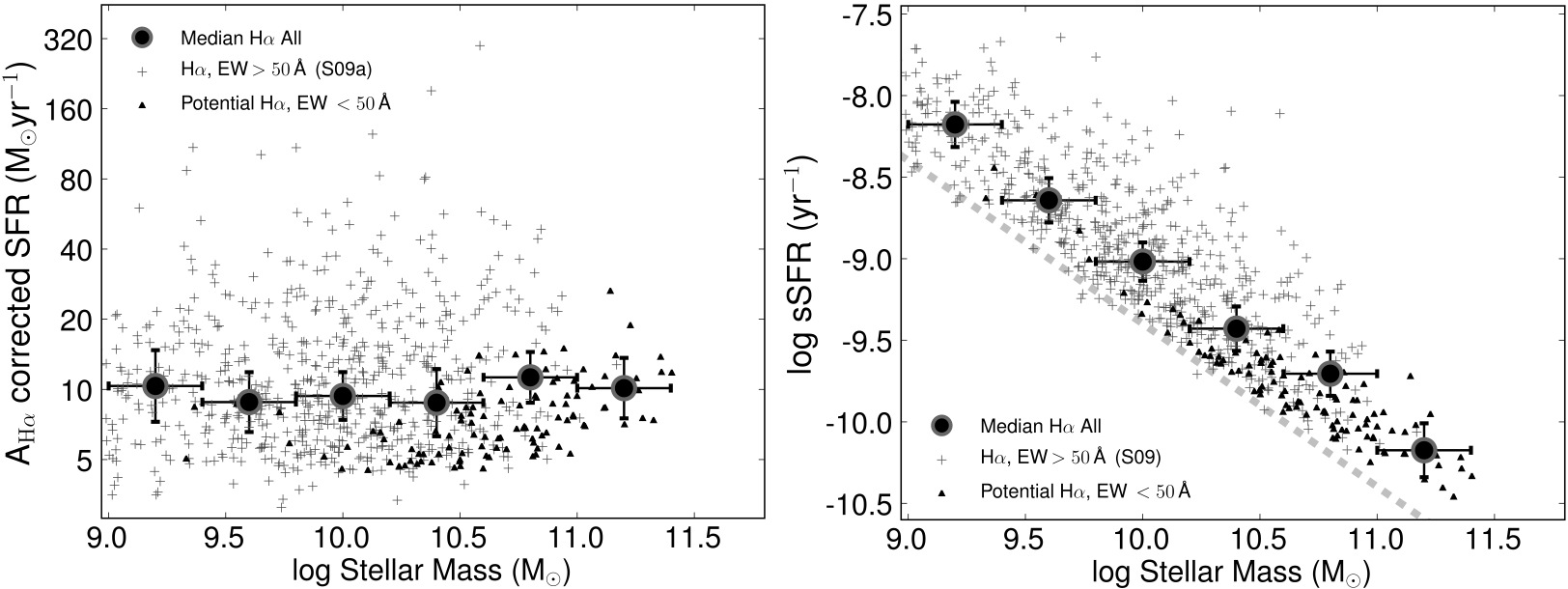}
\caption[wteta]{$Left$: The dust-extinction corrected SFR (H$\alpha$, SFRs obtained after applying the dust extinction from Garn et al. 2010) as a function of stellar mass for both the original S09a H$\alpha$ sample (EW$\,>50$\,\AA) and the possible H$\alpha$ emitters with EW$\,<50$\,\AA, showing that the median SFR is uncorrelated with stellar mass. Error bars present the 1\,$\sigma$ Poissonian errors ($\pm\sigma_m/\sqrt{\rm N}$). $Right$: the specific SFR (sSFR; SFR divided by stellar mass) falls with stellar mass, for both samples, although the steepness of such fall at the lowest masses is at least partially driven by the flux limit of the survey (dashed line; see discussion in main text). Both panels show very clearly how the nature of the samples with different EWs differs, with the S09a sample being made of galaxies with lower stellar masses, higher SFRs and thus much higher sSFRs at all masses, while the lower EW sample is mostly made of massive galaxies, presenting low sSFRs. \label{SFR_vs_mass}}
\end{figure*}

Figure \ref{dens_r0} shows how $r_0$ correlates with $\Sigma_{\rm c}$. The trend is found to be very well approximated (for $r_0$ in h$^{-1}$\,Mpc and $\Sigma_{\rm c}$ in Mpc$^{-2}$; see Figure \ref{dens_r0}) by $r_0\approx0.4\times\Sigma_{\rm c}+1.6$.

Having been able to find an empirical link between $\Sigma_{\rm c}$ and the correlation length of the H$\alpha$ emitters, one can associate different qualitative environments with local surface density (based on $r_0$). Studies such as \cite{Yang20005} found that galaxies in the poorest group environments typically have $r_0\approx6$\,h$^{-1}$\,Mpc, while the rich groups have $r_0\approx15$\,h$^{-1}$\,Mpc; for cluster environments, studies such as \cite{Mo96} find typical correlation lengths of $r_0>15$\,h$^{-1}$\,Mpc. For the purposes of the analysis, three main environments will be considered: field (including poor and rich field), poor groups and rich groups/clusters. Field environments are defined to be those with $\Sigma_{\rm c}<10$\,Mpc$^{-2}$ as these correspond to $r_0<6$\,h$^{-1}$\,Mpc. Poor group environments are considered to be those for which $\Sigma_{\rm c}=10-30$\,Mpc$^{-2}$, as samples with these densities yield the range of $r_0$ found for such groups ($6<r_0<15$\,h$^{-1}$\,Mpc). Rich groups/clusters are considered to be those at even higher densities, corresponding to $r_0>15$\,h$^{-1}$\,Mpc, for which $\Sigma_{\rm c}>30$\,Mpc$^{-2}$.

Of course, these simple classifications do not attempt to be fully accurate, nor do they aim to identify the exact transitions between what should be considered field, groups or cluster environments on their own. Nevertheless, they provide a valuable tool to associate the estimated local densities with different qualitative environments.

\subsection{Confirmed clusters and groups in COSMOS and UDS}

The on-sky distribution of H$\alpha$ emitters can be seen in Figure \ref{on_sky}, together with the different densities in which they reside in both the UDS (left panel) and the COSMOS (right panel) fields. By probing a very wide area, the H$\alpha$ survey is able to target a relatively wide range of densities. Particularly, by chance, the survey is able to probe a luminous and massive (M$_{500}\approx10^{14.3}$\,M$_\odot$) X-ray selected cluster in COSMOS at $z=0.835$ (see Figure \ref{on_sky}), along with other nearby lower luminosity X-ray groups at $z=0.85$ which have already been reported by the COSMOS team \citep{Finoguenov07,COSMOS_new_groupsXra}. These confirmed group/cluster regions are identified as high local projected density regions, with the calibration from \S2.7 clearly classifying them as group/cluster regions: this not only confirms that this study is able to probe a very wide range of environments, from the poorest fields up to rich groups/clusters, but it also backs-up the qualitative environment calibration used.

\cite{FinoguenovUDS} reports on a search for groups and clusters in the UDS field; their catalogue is used to confirm a reasonably rich X-ray group in the UDS field, along with 2 other lower luminosity selected groups within the narrow-band redshift coverage of H$\alpha$ (see Figure \ref{on_sky}). This further confirms the wide range of densities probed by the survey, not only in the COSMOS field, but also in the UDS field.

\section{Mass-dependences and downsizing} \label{down}

The complete sample of H$\alpha$ emitters at $z=0.84$ present typical stellar masses of M$=10^{10.1}$\,M$_{\odot}$, while the general population of galaxies at the same redshift are redder in the rest-frame U-V, and present a wider range of stellar masses.

As the left panel of Figure \ref{SFR_vs_mass} shows, the median H$\alpha$ star formation rate (corrected for dust extinction following Garn et al. 2010) for the entire sample does not correlate with stellar mass. The latter also reveals that the lower EW H$\alpha$ emitters are typically much more massive and less active. This is further confirmed by computing the specific star formation rates (sSFR, SFR per unit stellar mass) of the entire sample: as the right panel of Figure \ref{SFR_vs_mass} reveals, there is a significant decrease in sSFR with stellar mass for the two samples of H$\alpha$ emitters, with the lower EW H$\alpha$ emitters presenting the lowest sSFRs (stressing the importance of including these, since they allow for a much wider range of masses and sSFRs to be studied).

The observed sSFR-mass trend is similar to the one found by studies such as \cite{Feulner05_SSFR_MASS}, but is contrary to other determinations which suggest a much flatter relation between mass and sSFR \citep[e.g.][]{Daddi07,Dunne09,Rodighiero}. One reason for this relates to a selection bias of the current sample due to the fixed SFR limit (dashed line in Figure \ref{SFR_vs_mass}): at the lowest masses probed, low sSFR galaxies fall below the H$\alpha$ selection limit, biasing upwards the median sSFR determined. To estimate the mass below which this bias becomes significant, the survey limit of SFR\,$>3$\,M$_{\odot}$\,yr$^{-1}$ can be combined either with the typical median sSFR\,$\sim 10^{-9}$\,yr$^{-1}$ argued in the above papers, or with the 6 Gyr age of the Universe at $z\sim0.8$, by considering galaxies which form stars at a fairly constant rate across their cosmic history. These both suggest that a detection bias will set in below $\sim10^{10}$\,M$_{\odot}$ (a value confirmed by the analysis later in this section). At lower masses, the galaxies that are detected are mostly those with unusually high sSFRs (i.e., bursty).

This selection bias will therefore explain some or all of the steeper slope at low masses in Figure 3, but it cannot explain the strong trend at higher masses. Here, the difference from (and between) previous literature determinations largely arises from the definition of ``star-forming galaxies". In the current work, all galaxies with H$\alpha$ detections are classified as star-forming (SF) galaxies, leading to a clean selection cut in SFR. In contrast, previous works have often applied additional criteria to select the SF population. For example, Daddi et~al. analyse 24\,$\umu$m and ultra-violet data (i.e. broadly a SFR-selection, as in the current work) for a population of BzK-selected galaxies, but then they remove from consideration the galaxies classified as `passive BzKs' (a simple colour cut). This colour cut broadly corresponds to a cut in sSFR of their galaxy population, and has the effect of excluding the high-mass low-sSFR galaxies from their sample: it is not surprising that this then leads to a relatively flat mass-sSFR relation. Thus, the nature of the ``mass-sSFR relation for star-forming galaxies", which has been widely debated in the literature, is critically dependent on exactly which galaxies one considers to be star-forming.

\begin{figure}
\centering
\includegraphics[width=7.8cm]{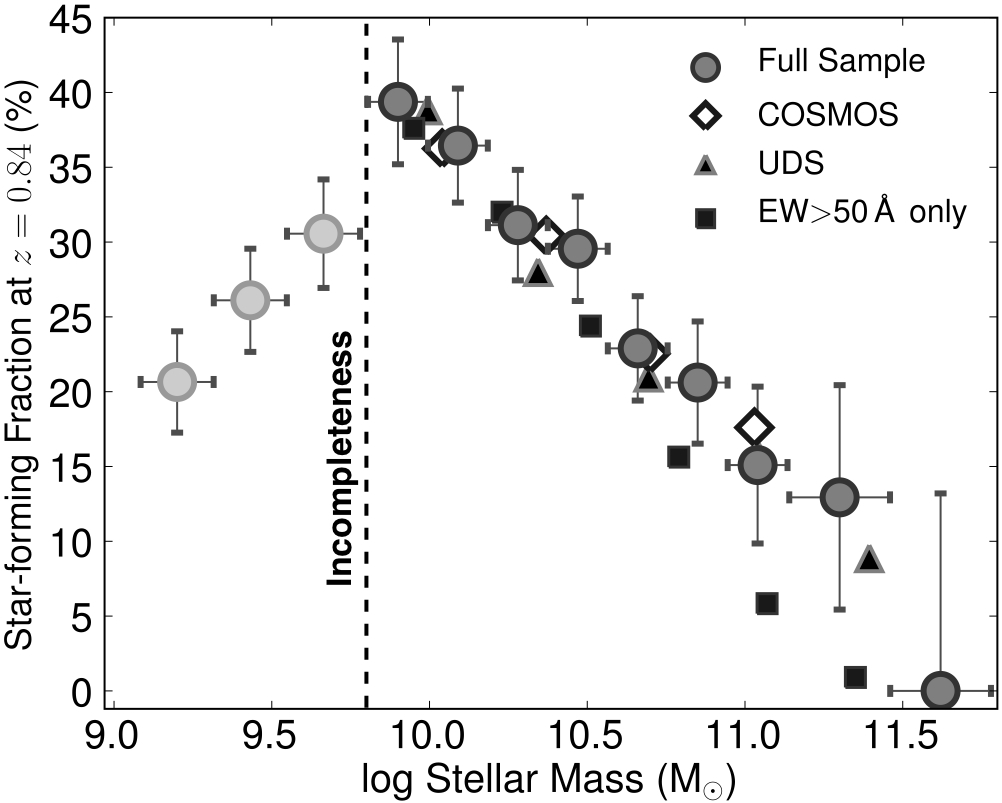}
\caption[wteta]{The fraction of galaxies forming stars with SFRs\,$>3$\,M$_{\odot}$\,yr$^{-1}$ when compared to the underlying population as a function of stellar mass for the entire sample, separately for the COSMOS and UDS fields, and for the S09a sample only. The figure shows that the fraction of star-forming galaxies with SFRs above the survey limit (SFR\,$>3$\,M$_{\odot}$\,yr$^{-1}$) is $\approx40$ per cent at $\sim10^{10}$\,M$_{\odot}$, but falls sharply to 0 as a function of stellar masses. The EW cut of S09a results in a sharper decline of the star-forming fraction, mostly because it misses the most massive galaxies with low specific star formation rates. Errors are Poissonian, but they also include a 20 per cent contribution from the contamination and completeness corrections applied. \label{downsi_COSMOS}}
\end{figure}

By comparing the sample of H$\alpha$ emitters (star-forming population) with the underlying galaxy population it is possible to determine the fraction of star-forming galaxies as a function of stellar mass, which can provide an important insight into where most of the star formation is occurring at $z\sim1$. Figure \ref{downsi_COSMOS} presents the results
(after correcting the underlying population for contamination and completeness as a function of mass; see Appendix A). It reveals that while for M\,$\approx10^{10}$ M$_{\odot}$ the fraction of star-forming galaxies is $\approx40$ per cent, it continuously drops as a function of increasing stellar mass, effectively reaching 0 per cent just below 10$^{11.5}$\,M$_{\odot}$; the same results are independently recovered for COSMOS and UDS. Using the S09a sample produces a sharper decline as a function of mass, mostly because the EW selection misses massive galaxies with very low sSFRs.

\begin{figure}
\centering
\includegraphics[width=7.8cm]{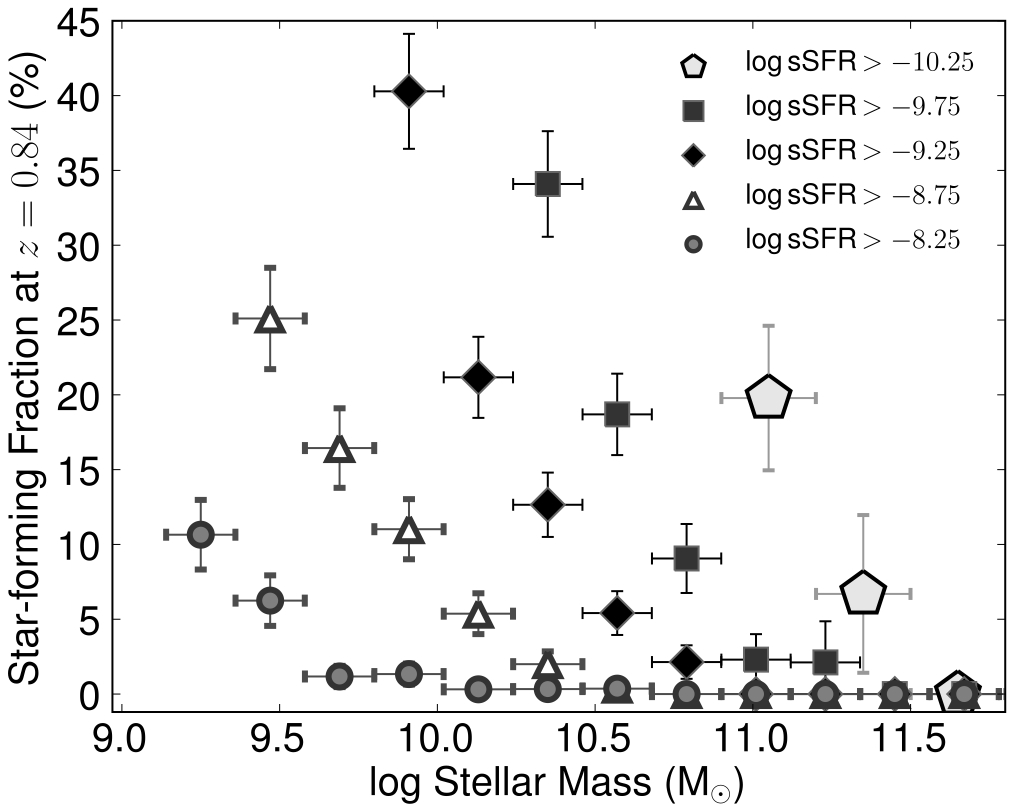}
\caption[wteta]{The fraction of star-forming galaxies as a function of stellar mass with a specific star formation rate greater than a given cut-off (corrected for completeness and contamination as a function of mass). This shows that the downsizing trend presented in Figure \ref{downsi_COSMOS} is recovered regardless of the sSFR cut-off used, and allows for lower masses to be probed. Errors are Poissonian and take into account the contamination and completeness from the photometric redshift selected population. Points are only plotted if above the SFR$>3$\,M$_{\odot}$\,yr$^{-1}$ survey limit. \label{downsi_SSFR_COSMOS}}
\end{figure}

Figure \ref{downsi_COSMOS} also reveals an apparent drop in the fraction of star-forming galaxies at lower masses ($<10^{9.8}$M$_{\odot}$). This is simply a result of the SFR limit in the H$\alpha$ sample that was discussed above\footnote{In order to properly probe star-forming galaxies with lower masses one requires a much deeper survey such as
ROLES \citep{Gilbank2010}.}. Nevertheless, specific star formation rate cuts can provide a view which broadly avoids this bias at the low mass end. Figure \ref{downsi_SSFR_COSMOS} presents the fraction of star-forming galaxies presenting sSFR above a certain value (points are only plotted if they are above the SFR limit of the survey, which avoids biases from the flux limit). The results reveal that the trend of a decreasing
star-forming fraction with increasing stellar mass is recovered for galaxies with a wide range of specific star formation rates, implying that the results are not simply biased by more passive-like, massive galaxies, dominating the very massive end of the star-forming fraction-mass relation. This analysis is also able to probe to lower masses, where the
same trend is recovered. Nevertheless, it should be noted that the trends recovered for different sSFR cuts are not simple linear functions and, particularly for high sSFR cuts, the fraction of star-forming galaxies above a certain stellar mass is effectively zero up to the highest masses probed.

These results indicate that mass-downsizing is already in place at $z\sim1$, with a sharp decrease of the star-forming fraction as a function of stellar mass for the highest masses. Therefore, the most massive galaxies at $z\sim1$ are mostly non-starforming, and even those that are forming stars present very low sSFRs.



\section{The Environment of H$\alpha$ emitters} \label{envirnmt}

\begin{figure*}
\centering
\includegraphics[width=16.2cm]{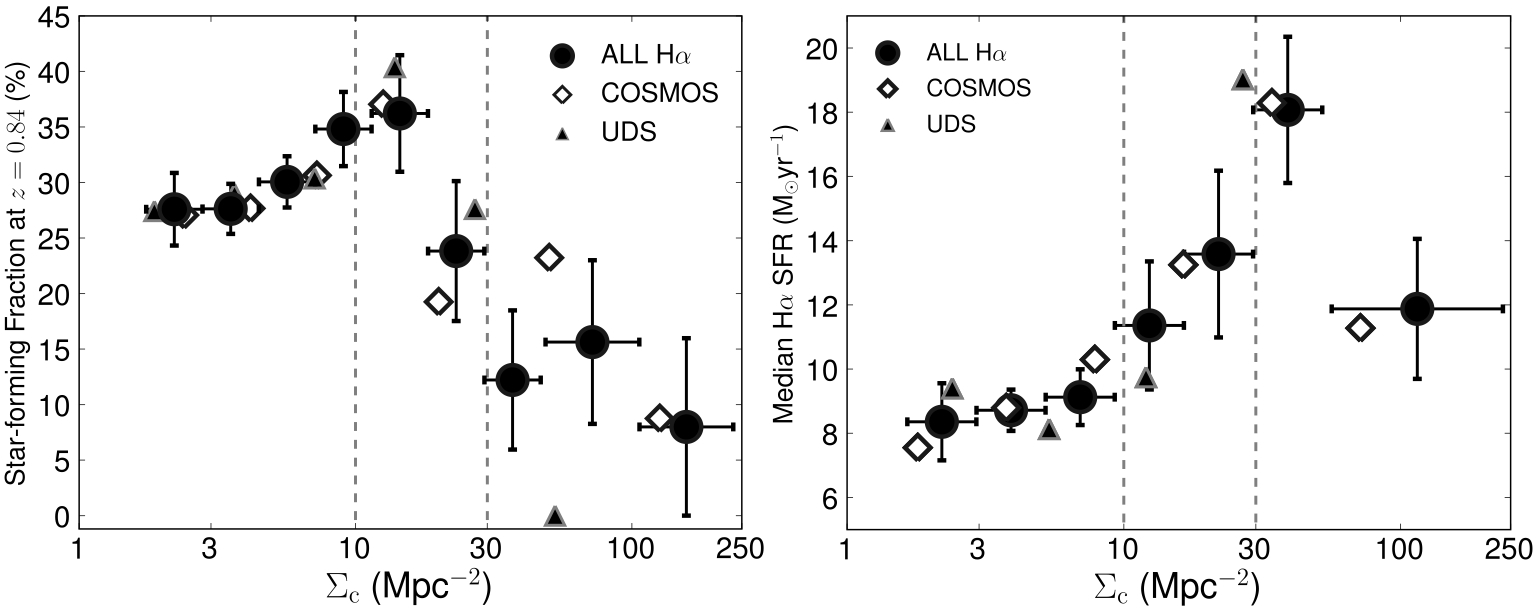}
\caption[wteta]{The fraction of the H$\alpha$-emitting galaxies when compared to the underlying population at the same redshift (left panel) and the median H$\alpha$ star formation rate of H$\alpha$ emitters (right panel), as a function of local projected galaxy surface density (for H$\alpha$ emitters with SFRs\,$>3$\,M$_{\odot}$\,yr$^{-1}$). The error bars present the Poissonian uncertainty for each bin and in the left panel also include a 20 per cent contribution from the completeness and contamination corrections applied. At the lowest densities, the star-forming fraction is relatively constant and only increases slightly, from $\sim30$ per cent at the lowest field densities probed to $\sim35$ per cent across the field environment. However, for higher densities, $\Sigma_{\rm c}>10$\,Mpc$^{-2}$, there is a steep decline of the star-forming fraction down to the highest (rich groups/cluster) densities probed, where the star-forming fraction is only $\approx5$ per cent. The results also show that the typical (median) star formation rate of H$\alpha$ emitters increases continuously from the lowest densities to $\Sigma_{\rm c}\sim$ 50 Mpc$^{-2}$, although it appears to fall again in the very richest environments. This corresponds to emitters residing in groups and cluster outskirts presenting typical SFRs which are about 2 times higher than those residing in the field. \label{environment}}
\end{figure*}

In this Section, the star-forming fraction is studied as a function of corrected surface density at $z=0.84$. The main result can be seen in the left panel of Figure \ref{environment}, which reveals that the star-forming fraction is relatively flat with density up to $\Sigma_{\rm c}\sim$\,10\,Mpc$^{-2}$ (with hints of a slight rise, but only at a $\approx1.5\sigma$ level), but then falls sharply for much higher densities. This seems to be associated with the qualitative transition between environments: the star-forming fraction (for galaxies producing stars at a rate higher than 3\,M$_{\odot}$\,yr$^{-1}$) decreases with increasing density once group regimes are reached, approaching zero for the highest densities; this is in good agreement with the results presented by \cite{Koyama09} at a similar redshift. The trend is the same in both COSMOS and UDS fields (even though the UDS field is not able to probe the highest densities found in the COSMOS X-ray cluster), revealing that this is not caused by a bias in one of the fields, nor due to the selection of the underlying population (this is very unlikely since the COSMOS and UDS photometric redshift catalogues are completely independent, obtained with different codes, and by different teams). It should be noted that the narrow-band filter is not a perfect top-hat and that some group/cluster members may have their H$\alpha$ line being detected towards the wings of the filter, affecting the likelihood of detecting them as star-forming. However, particularly because two independent fields are considered in the analysis, this effect is likely to even out in the statistics, with different groups detected at all points on the filter profile. Furthermore, as detailed in Appendix A, the trend is recovered regardless of the choice of underlying populations: only small differences are found once one applies the completeness \& contamination corrections given by Equation 1.

The right panel of Figure \ref{environment} shows that the median star formation rate of H$\alpha$ emitters rises with increasing local density, at least up to group densities, although there are hints that this relation does not hold for highest (rich groups/cluster) densities, consistent with previous results which find that, for rich environments, the typical SFR decreases with $\Sigma$ \citep[e.g.][]{Poggianti2008}. The results in this paper are also consistent with Sobral et al. (2010), which found that the correlation length of star-forming galaxies rises with H$\alpha$ luminosity. The relation between the median SFR and density is seen in both COSMOS and UDS, and is also observed for different H$\alpha$ luminosity limits adopted or when the analysis is restricted to spectroscopically confirmed H$\alpha$ emitters.

\begin{figure}
\centering
\includegraphics[width=8.2cm]{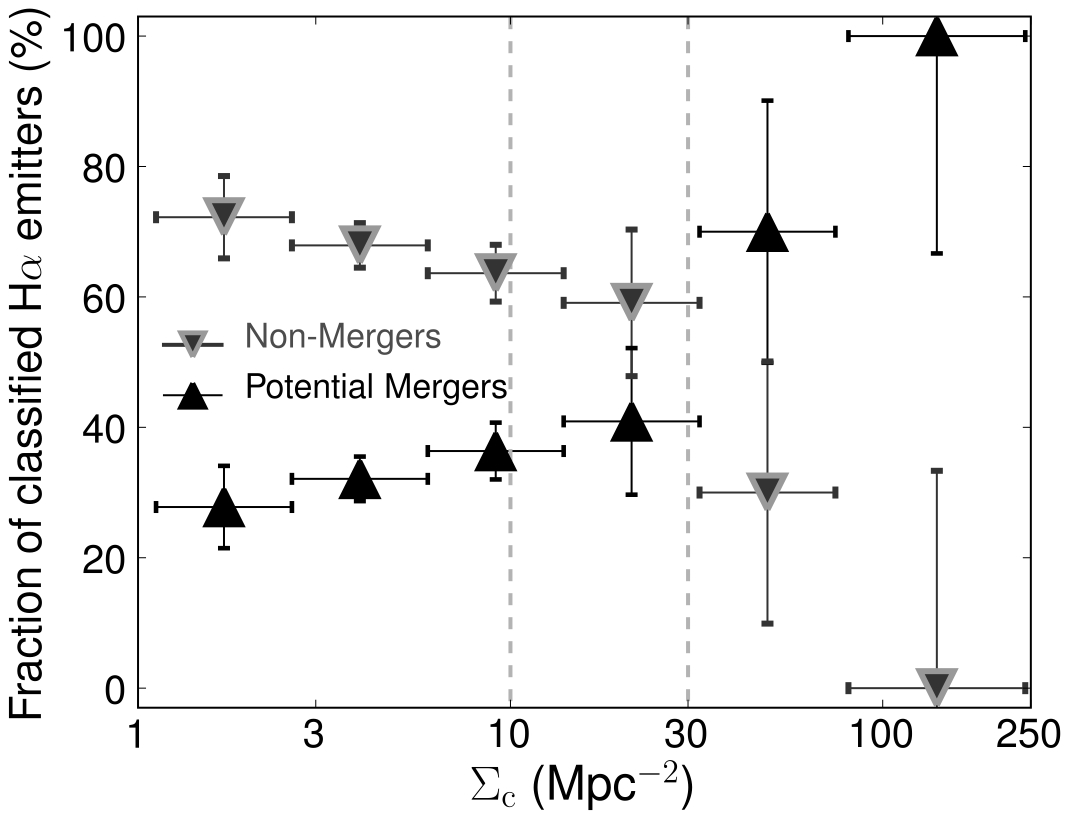}
\caption[wteta]{The potential merger and non-merger fractions as a function of projected local surface density (errors are Poissonian). This reveals that the fraction of potential mergers is reasonably flat at $\sim33$ per cent for field densities but is then found to increase continuously with density, reaching $\approx100$ per cent at rich groups/cluster densities. \label{merger_dens_fraction}}
\end{figure}

Moreover, as Figure \ref{merger_dens_fraction} shows, the connections between environment and star formation activity are likely to be related with the star formation mode: while star formation in the field is dominated by non-interacting galaxies, at rich group and cluster densities, the star formation seen at $z=0.84$ in H$\alpha$ is dominated by potential mergers; with this correlation being found at a 2.4$\sigma$ level. In fact, H$\alpha$ star-forming galaxies found in the X-ray core of the massive cluster in COSMOS are all mergers.

\begin{figure}
\centering
\includegraphics[width=7.8cm]{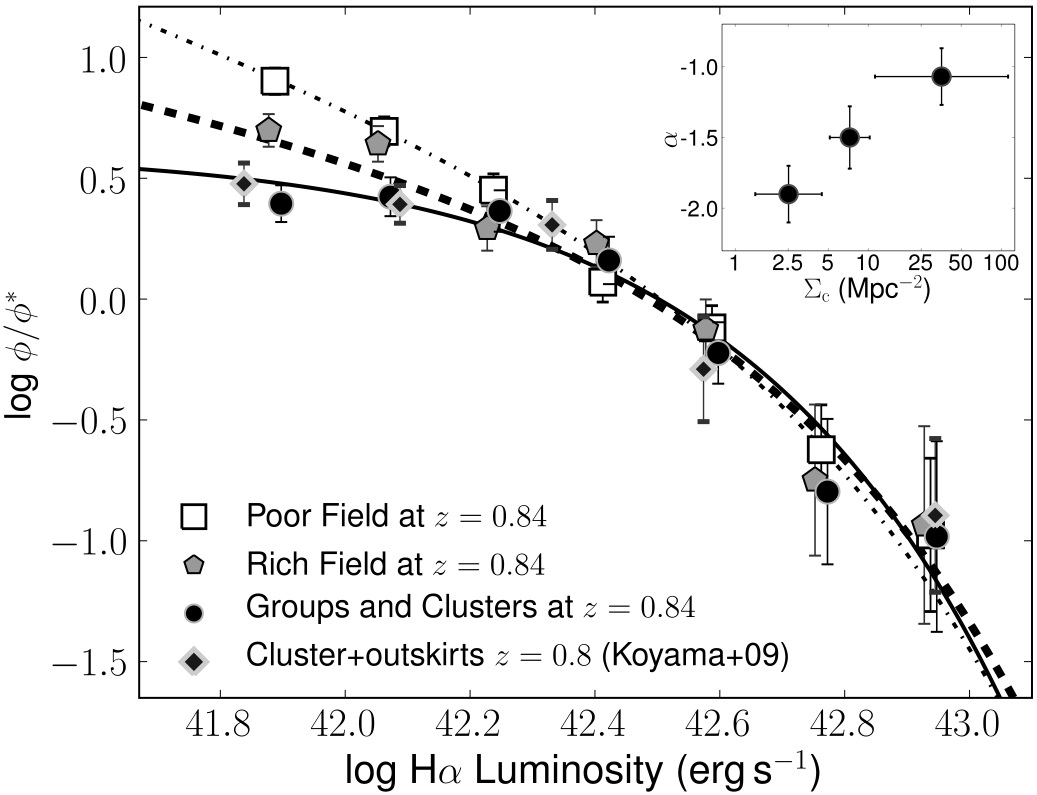}
\caption[wteta]{The normalised H$\alpha$ luminosity function for different environments, with the best-fit Schechter functions. While the characteristic luminosities, $L^*_{\rm H\alpha}$, of the different luminosity functions are in very good agreement within the errors, the results show a significant difference in the faint-end slope, $\alpha$, as a function of local surface density. As shown in the inset, this is found to be very steep ($\alpha\approx-1.9$) for the lowest densities (poor field, $\Sigma<5$\,Mpc$^{-2}$), shallower for medium densities (rich field, $5<\Sigma<10$\,Mpc$^{-2}$, with $\alpha\approx-1.5$) and very shallow for groups and clusters ($\Sigma>10$\,Mpc$^{-2}$), with $\alpha\approx-1.1$. Recent results from \cite{Koyama09}, which studied one cluster and its outskirts at $z=0.8$ (therefore equivalent to group and cluster densities) fully agree with the H$\alpha$ luminosity function derived for the $\approx$same local projected densities. \label{LF_environm}}
\end{figure}

The environment seems to be connected with a change in the star-forming fraction, the typical star formation rate and the nature of H$\alpha$ emitters at $z=0.84$; how are these changes manifested in the H$\alpha$ luminosity function? 
While it is nearly impossible to accurately determine the volume occupied by galaxies residing in different environments, one can arbitrarily normalise any obtained H$\alpha$ luminosity function to investigate changes in $L^*_{\rm H\alpha}$ and the faint-end slope, $\alpha$: this approach can provide the insight that one needs to further investigate the relation between star formation and environment.

Figure \ref{LF_environm} presents the normalised H$\alpha$ luminosity functions\footnote{H$\alpha$ luminosity functions are computed using the entire sample of H$\alpha$ emitters, which includes the low EW sample. Conclusions remain unchanged even if the low EW sample is excluded.} appropriately weighted as discussed in Section 2.3) for different local projected densities. By fixing $\phi^*$, both $L^*_{\rm H\alpha}$ and the faint-end slope, $\alpha$, are allowed to vary in order to obtain the best Schechter function fit to each luminosity function. The results reveal that while the value of $L^*_{\rm H\alpha}$ is consistent within the errors for all environments probed by the HiZELS survey at $z=0.84$ ($\log L^*_{\rm H\alpha}=42.47\pm0.05, 42.44\pm0.07, 42.45\pm0.09$\,erg\,s$^{-1}$ for poor fields, rich fields and groups and clusters, respectively), there is a strong difference in the steepness of the faint-end (see the inset of Figure \ref{LF_environm}). The faint-end of the LF becomes continuously shallower with increasing density, from $\alpha=-1.9\pm0.2$, for the lowest densities, to $\alpha=-1.5\pm0.2$ for rich fields and $\alpha\approx-1.1\pm0.2$ for groups and clusters. This clearly implies that the enhancement in the typical SFRs with environment directly results from the strong depletion of the faint-end and the shallow $\alpha$ in rich environments relative to the field, rather than an increase in the bright-end. This also reveals that the steepness of the faint-end of the H$\alpha$ LF in S09a is completely driven by field galaxies. The reader is referred to Section 6.2 for a discussion and possible interpretations of these results.

Figure \ref{LF_environm} also provides a direct comparison with results from \cite{Koyama09}; these authors studied a high density region at $z\sim0.8$, probing densities which are directly comparable with groups and cluster densities. The data are corrected for extinction using the relation presented in Garn et al. (2010) for a direct comparison. Koyama et al. found their data to be reasonably consistent within the errors with the total S09a LF (which contained all environments), but their data fully support the difference in the faint-end of the luminosity function between field and group/cluster environments found here, and completely agrees with the results derived for the same densities, supporting the results which show that the environment seems to be responsible for setting the faint-end slope of the LF, at least at $z=0.84$. A qualitatively similar result has been found for the 0024 super-cluster by Kodama et al. (2004) at $z=0.4$; however, while studies at even lower redhift \citep[see e.g.][]{Balogh02} have found different and mostly discrepant results regarding $\alpha$. It is clear that a comprehensive and large local study is still needed to really detail the role of the environment on the faint-end of the H$\alpha$ LF in the local Universe, to provide the low-redshift benchmark that can be used to look for any cosmic evolution..

\section{The nature of star formation and the mass-environment view}

\subsection{The star-forming fraction in the 2-D mass-environment space}

\begin{figure*}
\centering
\includegraphics[width=16.8cm]{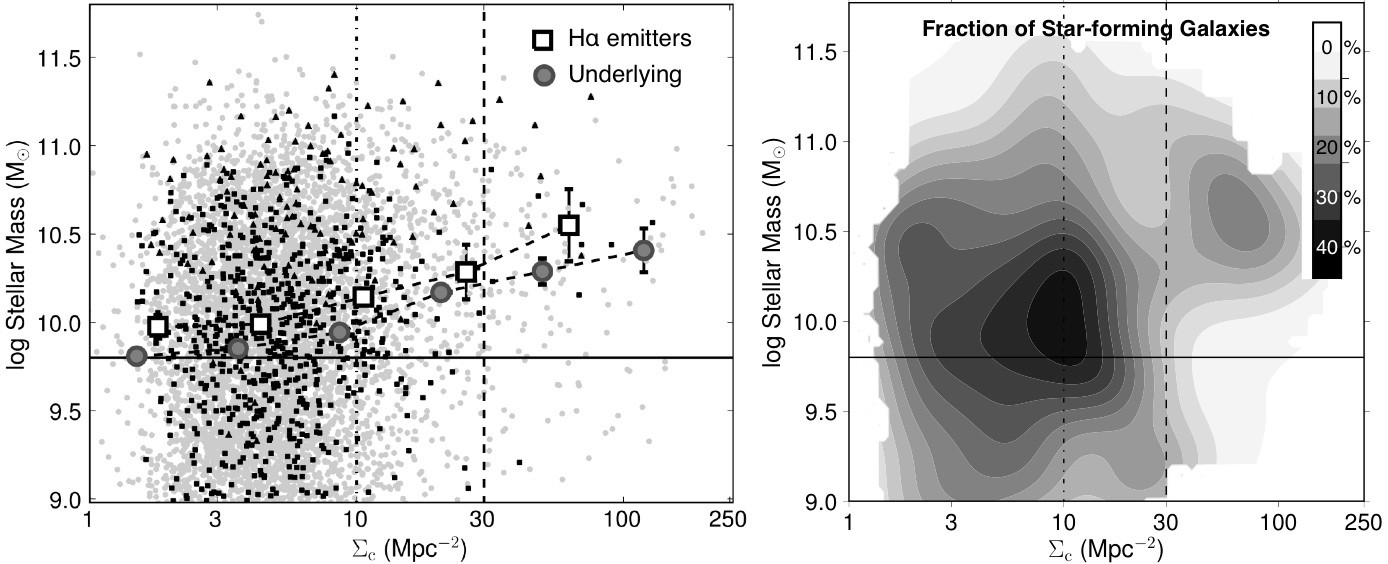}
\caption[wteta]{$Left$: The distribution of both H$\alpha$ emitters (filled squares: emitters with EW\,$>$\,50\,\AA; filled triangles: emitters with EW\,$<$\,50\,\AA) and galaxies within the underlying population (dots) in the 2D mass-environment plane. Vertical lines represent the qualitative transitions between field, groups and rich groups/cluster environments, while the filled horizontal line indicates the approximate mass for which the star-forming population becomes significantly incomplete due to its star formation rate limit (see Section 3). This reveals the expected correlation between median stellar mass and local density (larger symbols). Note that the median of the underlying population is evaluated over all galaxies, while the star-forming population is incomplete below the horizontal line, hence the apparently higher median mass for star-forming galaxies. $Right$: The fraction of galaxies that are forming stars as a function of both stellar mass and environment. This is calculated in regions of the 2-D space for which it is possible to gather 5 or more background galaxies in a (Gaussian) smoothing radius up to 1$\sigma$ of the typical errors in mass and environment -- regions with less than 5 background galaxies are not shown. The latter reveals that the overall behaviour of the star-forming fraction as a function of stellar mass or environment depends, to some extent, on the environment or stellar mass, respectively. \label{SFF_mass_environment_2d}}
\end{figure*}

It has been shown that star formation has strong dependences on both mass and environment at $z\sim1$. Nevertheless, with environment and stellar mass expected to be somewhat correlated, how related are the dependences of star formation on mass and environment and are these intrinsically connected? Figure \ref{SFF_mass_environment_2d} presents the mass-environment 2-D space for H$\alpha$ emitters and for the entire underlying population. This reveals the expected trend of an increase in the typical stellar mass with increasing local projected density for both populations of galaxies (with a change of $\approx0.5$ dex in the median mass from low to high densities).

To attempt to distinguish any potential inter-dependence of the roles of mass and environment on star formation at $z\sim1$, the analysis presented in the previous 2 sections is repeated in the 2-D space of mass and $\Sigma_c$. The right panel of Figure \ref{SFF_mass_environment_2d} presents the star-forming fraction evaluated within this space, in regions where this can be calculated (see caption for details). While this clearly illustrates the overall behaviour which has already been presented, the fine details suggest significant connections between the dependence of the star-forming fraction on mass and environment. In particular, the results show that the weak increase of the star-forming fraction with density within the field regime and its subsequent sharp decline at higher densities is predominantly driven by galaxies of typical mass ($\approx10^{10}-10^{10.5}$\,M$_{\odot}$) in the sample (please note that, as described in Section 3, there is a significant incompleteness effect at the lowest masses; this is indicated by the horizontal filled line in the Figures). Interestingly, whilst the mass-downsizing trend is driven by galaxies in the dominant field and group environments, the results reveal an increase in the fraction of star-forming galaxies with density at the highest masses and, equivalently, an apparent reversal of the mass-downsizing trend in the richest environments.

\begin{figure*}
\centering
\includegraphics[width=15.8cm]{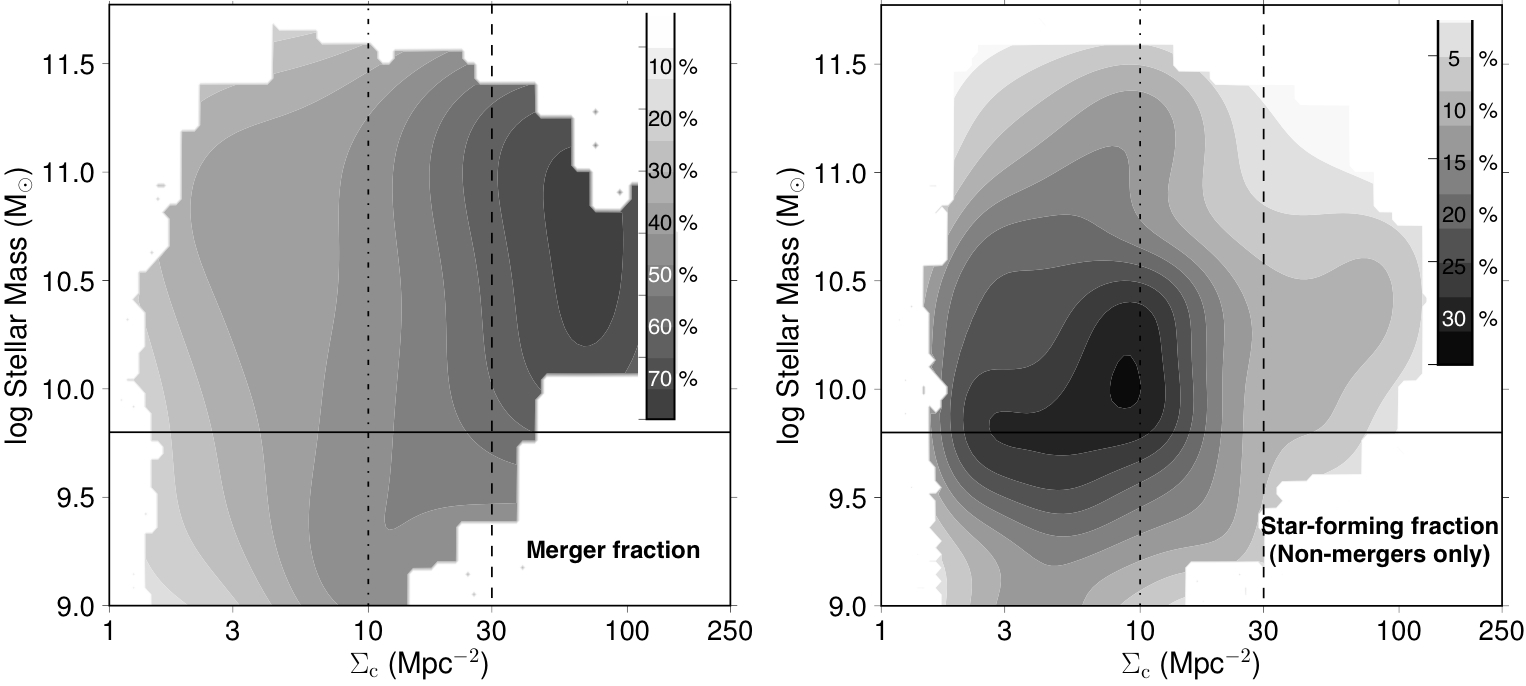}
\caption[wteta]{$Left$: The fraction of potential mergers within the star-forming population of galaxies (in COSMOS). This reveals that star formation occurring at the densest environments is potentially merger-driven. $Right$: The non-merger driven star-forming fraction, revealing that the trends with environment and mass become even stronger once potential mergers are removed from the sample. This is calculated in regions of the 2-D space (COSMOS only) for which it is possible to gather 5 or more background galaxies in a (Gaussian) smoothing radius up to 1$\sigma$ of the typical errors in mass and environment -- regions with less than 5 background galaxies are not shown. \label{2D_merg_quiescent}}
\end{figure*}

Earlier, it was shown that the fraction of potential mergers increases with local density. To investigate whether this change in the potential nature of star formation can shed some light onto understanding what is happening at relatively high masses and high densities, the left panel of Figure \ref{2D_merg_quiescent} presents how the fraction of star-forming galaxies that are classified as potential mergers (in the COSMOS field) varies within the 2D mass-environment space. It clearly reveals a strong increase in the fraction of potential mergers with environment, and, particularly, that star formation at the highest densities is mostly driven by potential mergers, regardless of mass. It is therefore clear that the behaviour of the star-forming fraction within the studied 2D space reflects the role of potentially merger-driven star formation for very high densities, where merging events are more likely to happen. This is shown more clearly in the right panel of Figure \ref{2D_merg_quiescent}, which repeats the analysis of Figure 9, but now restricted to isolated star-forming galaxies only (i.e. excluding all potential mergers). It is found that the fraction of non-merging star-forming galaxies peaks roughly at the mass completeness limit (for the SFR limit of the current sample, $\approx10^{9.8}$\,M$_{\odot}$), and in the transition between fields and groups, but shows strong general declining trends with both increasing mass for all environments and with increasing local density for all masses. This reveals that non-merger driven star formation is strongly suppressed in both rich groups/cluster environments and for high stellar masses, implying that once potential mergers are neglected, stellar mass and environment both play separate and important roles.

\subsection{The mass-SFR-environment dependence}

Section 4 showed that the typical SFR increased with density at least up to groups/cluster outskirts densities. Here this is investigated further, by examining how the median star-formation rates vary with environment for sub-samples of star-forming galaxies with different stellar masses. Figure 11 reveals that while a strong SFR-environment trend is seen in low to moderate stellar mass galaxies, it does not hold for the most massive galaxies (M\,$> 10^{10.6}$\,M$_{\odot}$), for which the median SFR seems to remain unchanged with increasing local density. Since the typical stellar mass of galaxies increases with density, in terms of sSFRs one finds that for the most massive galaxies, the
median sSFR decreases with increasing local density (at a 2$\sigma$ significance level), whilst this trend is continually reversed with decreasing stellar mass, becoming a significant positive correlation (3.5$\sigma$) at the lowest masses probed. It should be noted that although the most pronounced difference in the SFR-$\Sigma$ behaviour is in the densest environments, where the merger fraction is high, the same trends are recovered (albeit weakened by $\sim 0.6 \sigma$) if the analysis is conducted using only non-merger star-forming galaxies. Therefore, merging activity by itself is not enough to explain the enhancement of SFR in low-to-medium mass star-forming galaxies. These results imply that there is some inter-dependence in the influence of mass and environment on galaxy evolution.

\begin{figure}
\centering
\includegraphics[width=7.8cm]{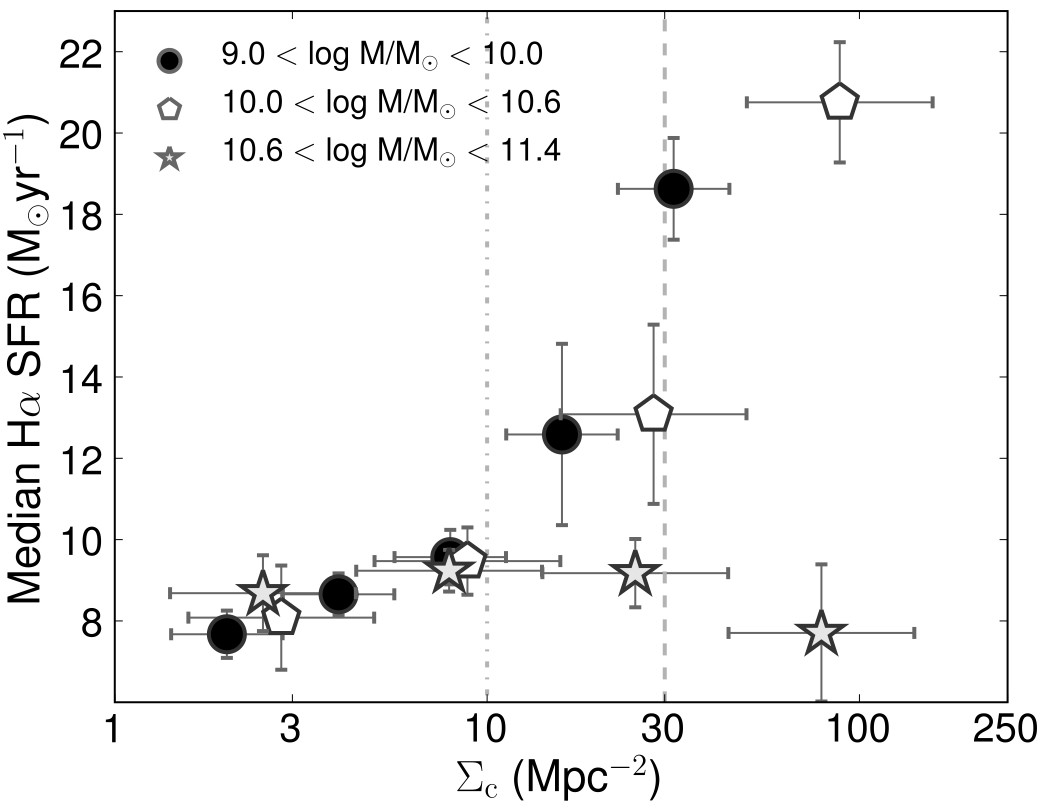}
\caption[wteta]{The dependence of the median SFR on environment ($\Sigma_c$) for star-forming galaxies with different stellar masses. Error bars present the 1\,$\sigma$ Poissonian uncertainty ($\pm\sigma_m/\sqrt{\rm N}$). The results show that while the increase of the median H$\alpha$ SFR with environment is valid for galaxies with stellar masses in the range $\approx10^{9.0-10.6}$\,M$_{\odot}$, the most massive galaxies in the sample (stellar masses $>10^{10.6}$\,M$_{\odot}$) do not have their typical SFR enhanced with environment. \label{SSFR_environment}}
\end{figure}

\subsection{The colours of star-forming galaxies in the 2-D mass-environment space}

\begin{figure*}
\centering
\includegraphics[width=15.8cm]{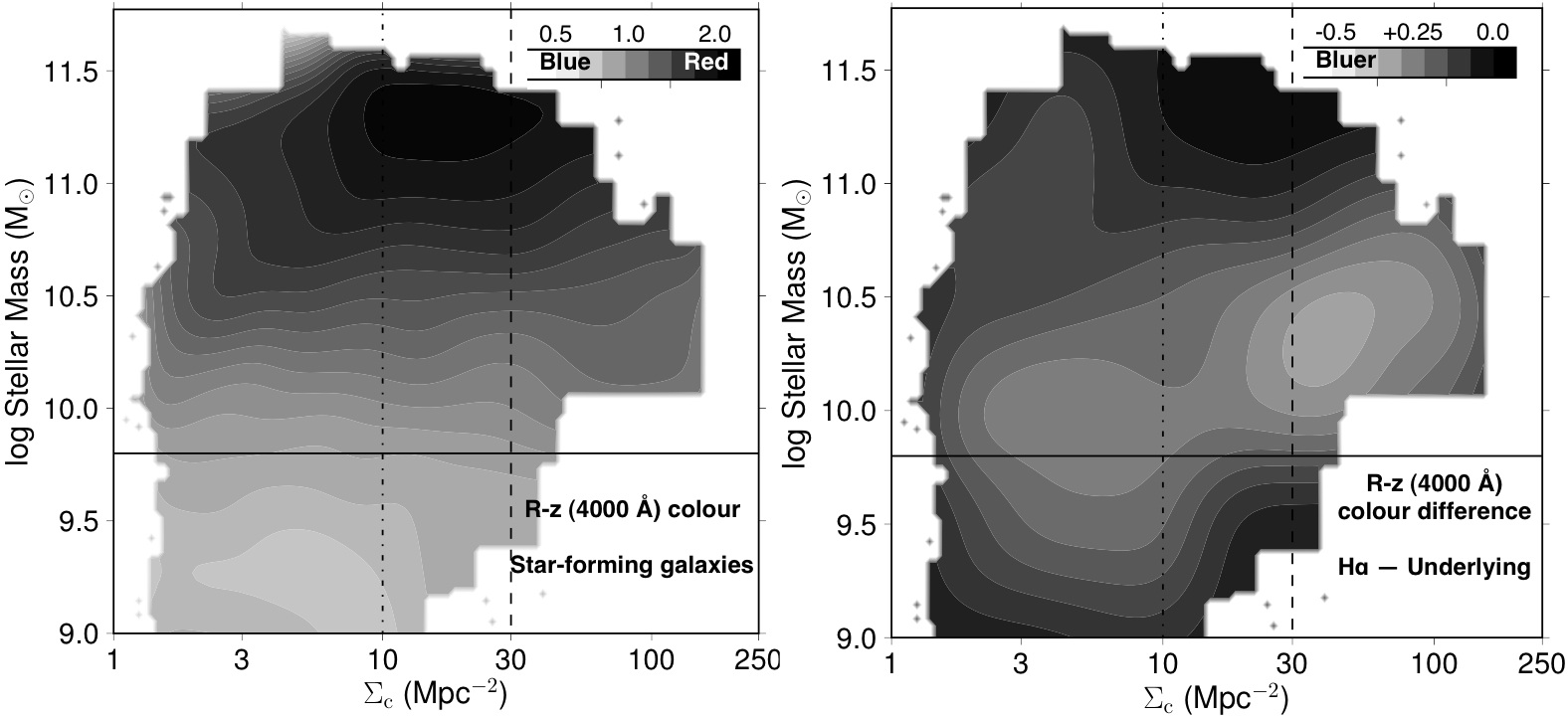}
\caption[wteta]{$Left$: The distribution of the median R-z colour (probing the 4000\,\AA \, break at $z=0.84$) within the mass-density 2D space for the H$\alpha$ star-forming galaxies, revealing that stellar mass is the main colour predictor as the environment only correlates weakly with colour. $Right$: The difference in the median R-z colour between the population of star-forming galaxies and the complete underlying population, revealing that although star-forming galaxies are consistently bluer than the underlying population, the colour difference is a clear function of mass and also depends on the environment. \label{2Dcolour_2}}
\end{figure*}

The nature of the H$\alpha$ emitters occupying different regions in the 2D space being probed can be made even clearer by looking at the colours of emitters. For $z\sim0.8$, this can be simply done using $(R-z)$ because these span the 4000\,\AA \, break, and are sufficiently close so that dust extinction is unlikely to be a significant problem. Figure \ref{2Dcolour_2} (left panel) presents how the median colours of the H$\alpha$ emitters vary across environment for different stellar masses. The results suggest that stellar mass is the most important quantity for determining the 4000\,\AA \, break colour (and whether the H$\alpha$ emitter is consistent with already being in the red sequence or still occupying the blue cloud) of star-forming galaxies. Figure \ref{2Dcolour_2} also reveals that the environment plays a much weaker role for determining the colour of star-forming galaxies, even though the reddest H$\alpha$ emitters are found in group environments (but those are also very massive). The 2-D colour distribution is found to be qualitatively the same for both potential mergers and non-merging galaxies, showing the same trends, although potential mergers are consistently bluer (by $\approx0.2$ in colour) than the non-mergers throughout the 2-D space studied.

Further information can be obtained by comparing the colour distribution of the star-forming galaxies with that of the underlying population. As the right panel of Figure \ref{2Dcolour_2} shows, the star-forming population is bluer than the underlying population at all masses and environments. However, the results indicate that even though the main colour predictor for star-forming galaxies is stellar mass, the difference in colour between the star-forming population and the underlying population of galaxies depends on both mass and environment. 
Particularly, the colour difference is greatest at lower masses (above the mass completion limit -- see Section 3), where the underlying population contains a mix of blue and red galaxies, while the star-forming population has very few red ones, thus leading to the difference. However, such difference is negligible at the very highest masses, where even star-forming galaxies are very red. Moreover, the fraction of redder galaxies at moderate masses is found to be higher in denser environments for the underlying population (the underlying galaxy population has a weak trend of colour with increasing density), but the star-forming population has essentially no trend with environment for a fixed mass (see left panel of Figure \ref{2Dcolour_2}), thus resulting in the environmental trend of the colour difference.


\section{Discussion}

It is clear that the bulk of the most massive galaxies that are already in place at $z=0.84$ are non-star-forming (star formation rates lower than the HiZELS limit, 3\,M$_{\odot}$\,yr$^{-1}$) and many of those which are star-forming have low sSFRs and have stronger 4000\,\AA \, breaks (are redder) than lower mass galaxies. These results clearly point towards mass downsizing already being in place for the general population of star-forming galaxies studied at $z\sim1$, although potential merging activity (which becomes more likely at higher densities) can be responsible for triggering some star formation even in massive, apparently red-sequence galaxies.

In addition to this, the environment has a significant role in the nature and evolution of star-forming galaxies at $z\sim1$. For local surface densities associated with the field regime, and for moderate stellar masses, the fraction of star-forming galaxies is relatively flat with increasing environmental density, while the median star formation rate of H$\alpha$ emitters is found to increase with density. However, for higher densities, the star-forming fraction falls sharply towards $\sim0$ per cent for the highest rich group/cluster densities. The environment is also found to play an important role in setting the faint-end slope of the H$\alpha$ luminosity function, with the results showing that under-dense regions have H$\alpha$ LFs with very steep faint-end slopes, while the highest densities have a shallow faint-end slope -- this confirms analytic predictions such as those presented by \cite{Trenti_Stiavelli}. If this holds at different epochs (particularly, at higher redshift), then it suggests that the steepening of the faint-end slope with redshift that has been argued by many authors \citep[for both H$\alpha$ and UV selected samples, e.g.][]{Hopkins2000,Bouwens2007,Reddy2009}, may be mostly a consequence of structure formation, since at high redshift groups and clusters of galaxies become rarer and rarer and the field environment dominates. At lower redshift, on the other hand, the discrepancies in the faint-end slope discussed in S09a are likely to be caused by cosmic variance, with different surveys targeting regions with different densities.

When one combines the dependences of star formation on mass and environment, a sharper picture arises: this shows that the increase in the star-forming fraction in the field regime is dominated by median mass ($\approx10^{10}-10^{10.5}$\,M$_\odot$), essentially non-merging galaxies, and that the bulk of star-forming galaxies in high density environments are red, massive galaxies undergoing a merger episode. It is also found that the fraction of non-merging star-forming galaxies strongly declines in rich groups/cluster densities for all masses, and for high masses for all environments, implying that both mass and environment independently suppress non-merging star-formation. However, this also shows that the enhancement of the typical SFRs with increasing density is only happening at $z\sim1$ for galaxies with masses lower than $10^{10.6}$\,M$_{\odot}$ -- for the most massive star-forming galaxies the environment does not seem to alter the typical SFRs and their properties seem to be mostly set by their stellar mass.

\subsection{Comparison with other studies}

Studies at $z\sim1$ and above have found apparent discrepant results for the relation between star formation and environment, with some authors claiming to have found an inversion in that relation when compared to the local Universe. In particular, \cite{elbaz} study the star formation rate-density relation at $z\sim1$ and find that, instead of declining, SFRs of star-forming galaxies are enhanced with local projected density; \cite{cooper} find similar results. These studies typically probe field to poor group environments. On the other hand, \cite{Poggianti2008} presents a detailed study of a large sample of relatively high redshift groups and clusters (up to $z\sim1$). Contrarily to \cite{elbaz} and \cite{cooper}, \cite{Poggianti2008} find that SFRs decrease with environment in the high density regime (but the results agree with a rise in the mean SFRs -- which are [OII] based -- with density up to $\Sigma\sim30$\,Mpc$^{-2}$, and with the fall after that). 

\cite{Patel} studied a $z\sim0.8$ cluster and the surrounding regions (using a mass limited sample, but selecting only relatively massive galaxies, and making use of MIPS 24\,$\umu$m data) to show that the specific star formation rate decreases with increasing density at $z\sim1$ and argued that star formation is suppressed with density even at that epoch. The authors comment on the big discrepancy of their results when compared to \cite{cooper} or \cite{elbaz}, and, whilst they argue that the differences could be arising from a combination of different effects, they are not able to reach a definitive answer. \cite{Feruglio} also study the role of the environment on star-formation and, using a sample of luminous and ultra-luminous infrared galaxies (LIRGs and ULIRGs) in COSMOS, find that for such massive galaxies there is not an increase in SFR with density up to $z\sim1$.

Fortunately, the large sample and the wide range of environments and stellar masses within the HiZELS sample provides an ideal opportunity to attempt to reconcile these results. Indeed, one finds, just like \cite{elbaz}, that the environment enhances SFRs at least up to groups/cluster outskirts environments, but then decreases at higher densities as found by Poggianti et al. Furthermore, if one looks at how the sSFR of massive galaxies (comparable to those studied by Patel et al. 2009) varies with environment, the same trend is recovered: sSFRs of massive galaxies decrease with $\Sigma$, mostly because median SFRs of massive galaxies are largely insensitive to environment, but the median mass increases at higher densities, leading to a lower median sSFR. Note that such a trend is reversed for lower mass galaxies, where the increase in SFR with density is fully valid; this reveals that the results from Patel et al. (a mass-selected sample), or \cite{Feruglio} (a sample of massive LIRGs and ULIRGs) are highly dependent on the stellar mass selection, and that these do not agree with results from \cite{elbaz} mostly because galaxies in the later study are selected from their star formation emission lines (therefore bluer, less massive galaxies).

Furthermore, \cite{Poggianti2008} also finds that the fraction of star-forming galaxies falls continuously for densities significantly larger than $\Sigma\sim10$\,Mpc$^{-2}$ (the study is not sensitive to field densities). Similar results are recovered by Koyama et al. (2010), who studied a cluster at $z=0.8$ and found that the star-forming fraction (H$\alpha$ emitters) falls from groups to the cluster core, while \cite{Feruglio} finds that the fraction of LIRGs and ULIRGs also decreases with increasing density up to $z\sim1$. These are the same trends found in this paper, but the current work extends this to much lower densities and, overall, probes a wider range of densities.

Together, the results in this paper are able to simultaneously agree with previously apparently contradictory results, revealing that such discrepancies are caused by the studies focusing on galaxies with significantly different masses and, most importantly, residing in different environments.

The analysis presented in this paper contributes towards improving our understanding of galaxy evolution and the key epochs for significant transitions. While the median SFR is enhanced with density up to moderate densities, for a star-forming selected population, it has also been shown that the behaviour of that relation depends on stellar mass. Furthermore, while merging does become important at the highest densities, it is also found that the SFR-$\Sigma$ relation is still recovered even when considering non-mergers only. This surely implies that the physical processes through which the environment enhances the typical SFRs cannot be fully explained by (major) mergers. Therefore, galaxy harassment, or the acceleration of the intergalactic gas infall in rich groups and cluster outskirts are likely to play a significant role in enhancing star-formation at high densities. Since these processes are more likely to affect galaxies with higher gas contents, it may be understandable that the trends are seen most strongly at low and medium masses.

Furthermore, the star-forming fraction drops from fields to groups and clusters, revealing no qualitative evolution from the local Universe: very rich environments already have a low star-forming fraction even at $z=0.84$ (and the star formation activity occurring at high densities is potentially merger-dominated). Particularly when one excludes potential mergers, the environmental trend of a declining star-forming fraction for high densities is recovered at all masses, revealing that the environment plays an important role which is roughly independent from that of stellar mass. This is in good agreement with the results presented by \cite{Cooper10}, at a similar redshift, and qualitatively similar to the clear independence of the quenching roles of stellar mass and environment found by \cite{Peng} using SDSS. Since HiZELS is able to agree with previous results and provide a much broader view, it further confirms that the densest regions at $z\sim1$ are already the sites where one finds predominantly red sequence galaxies and low star-forming fractions, as found in previous cluster studies. However, with studies such as \cite{Hayashi09} showing that the star-forming fraction does not drop with increasing density even for the densest regions at $z\sim1.5$ (but that some of [OII] emitters within the cluster core are already in the red sequence), it seems likely that the latter stages of the environment transformation happen between $z\sim0.8$, where one finds red H$\alpha$ emitters at cluster cores, but a fraction of star-forming galaxies which already declines steeply, and $z\sim1.5$, where such fraction seems to hold up to the highest densities; the same view seems to be recovered by using 24\,$\umu$m data \citep{Hilton_XMM_24um}.

\subsection{The role of the star formation mode: potential mergers vs. non-mergers} \label{Ha_LF_environmt}

S09a find that non-merging galaxies (mostly disc galaxies) dominate the star formation rate density of the Universe at $z\sim0.8$, and it is now clear that the bulk activity of such galaxies is occurring in the field and poor groups.

Nevertheless, while the fraction of non-merger driven star-forming galaxies falls sharply at high densities, there is still some star formation found in the richest environments. The latter is merger dominated, and this is linked with the increase in the median SFRs of H$\alpha$ emitters and, to some extent, with the fall of the star-forming fraction with environment.

Indeed, in group and cluster environments, $\sim80$ per cent of star-forming galaxies are potential mergers. This is somewhat expected, as mergers are more likely to happen in dense environments. However, merger-driven star formation at $z=0.84$ is likely to result in starbursts which quickly exhaust the available gas (which, for the most massive galaxies that are already reasonably red, is likely to be low), and are expected to produce elliptical galaxies. This would then further drive the decline of the overall star-forming fraction with environment and would be consistent with a significant population of mergers at high densities at slightly higher redshifts, as argued by S09a to explain the evolution of the H$\alpha$ luminosity function. Thus, as the densest regions increase the merger chances, they can lead to a much quicker migration of blue cloud galaxies into the red sequence. This would then establish a higher fraction of passive galaxies in the densest regions much earlier than in the field, where potential merging events are not dominant.

At higher redshift ($z=1.2$), \cite{Ideue} presented an environment analysis using [O{\sc II}]\,3727 emitters selected from a narrow-band survey, accessing the potential role of interacting galaxies as a function of environment in the COSMOS field (but instead of morphologically classifying the galaxies, they looked at the fraction of isolated against non-isolated galaxies). The authors find an increase in the fraction of non-isolated galaxies with density. However, while this is fully consistent with the results presented in this paper and with the interpretation, it should be noted that \cite{Ideue} define non-isolated galaxies as galaxies separated by distances of less than 80 kpc. Whilst this is not a robust criteria for identifying a merger, it also strongly correlates with the local density estimate, even if the merger fraction does not go up with density; therefore, an increase with density may be expected as a bias, regardless of the nature of the galaxies (real mergers or non-mergers).

The results of this paper can also be used to interpret the influence of the environment on the H$\alpha$ luminosity function. In S09a the H$\alpha$ luminosity function derived for just potentially merging sources was found to have a shallow faint-end slope, $\alpha$. With this study it seems likely that the low $\alpha$ is a consequence of a higher density environment: galaxies in high densities will increase their merging chances and reside in potentially gas-rich regions at $z\sim1$, making it hard to maintain low, isolated star formation rates.

The results imply that higher densities enhance star formation in star-forming galaxies and that seems to be valid at all cosmic times; that's why denser regions have the most massive galaxies; however, because star formation has been enhanced, galaxies form stars quicker and sooner, and this leads to the mass downsizing observed since $z=1$ down to the local Universe.

\section{Conclusions}

The conclusions of this study can be summarised as follows:

\begin{itemize}

\item The well-defined HiZELS sample of H$\alpha$ emitters at $z=0.845\pm0.015$ down to SFRs\,$>3$\,M$_{\odot}$yr$^{-1}$ covering a large sky area has been used to conduct a detailed study of the dependence of star formation on mass and environment. This is able to reconcile previous results and provide an integrated, clearer view of galaxy evolution at $z\sim1$, particularly by taking advantage of the excellent underlying data in the COSMOS and UDS fields, but also because the survey probes significantly massive, rich groups/clusters, providing a truly panoramic H$\alpha$ survey over a very wide range of environments.

\item The star-forming fraction is found to strongly decline with stellar mass, from $\approx40$ per cent for 10$^{10}$\,M$_{\odot}$ galaxies to effectively zero above 10$^{11.5}$\,M$_{\odot}$. Specific star-formation rates also decline continuously with stellar mass. Mass downsizing is therefore fully in place at $z=0.84$.

\item The median SFRs of H$\alpha$ emitters at $z=0.84$ increases as a function of local surface density for both field and group environments, but the trend is stopped for the highest densities, where it is likely to fall. This trend is driven by star-forming galaxies with stellar masses up to $\approx10^{10.6}$\,M$_{\odot}$; higher-mass galaxies seem to have median SFRs which are mostly independent of the environment they reside in. With this result being recovered even when one excludes potential merger-driven star-formation, it is likely that at least part of the environmental enhancement must come from other physical processes, such as galaxy harassment, or the acceleration of intergalactic gas infall in groups and cluster outskirts.

\item The fraction of galaxies that are forming stars is relatively flat with density within the field regime, but there is a clear, continuous decline once group densities are reached, resulting in the fall of the star-forming fraction from fields to rich groups/clusters, as seen in the nearby Universe and in studies focusing on the highest densities at $z\sim1$.

\item The environment is responsible for shaping the faint-end of the H$\alpha$ luminosity function: the faint-end slope, $\alpha$ is found to be very steep ($\alpha=-1.9\pm0.2$) for the poorest regions, becoming shallower ($\alpha=1.5\pm0.2$) for rich fields and even shallower ($\alpha=-1.1\pm0.2$) for groups and clusters.

\item Star formation at the highest densities is dominated by potential mergers, with likely implications on the nature and duration of the star formation activity, and suggesting that at least part of the environment quenching of star formation seen in the local Universe is driven by merging events at higher redshift. Neglecting potential merger-driven star formation results in stronger and mostly independent trends of a declining fraction of star-forming galaxies with both increasing mass and environment.

\item The 4000\,\AA \, colour of star-forming galaxies correlates strongly with stellar mass, but only weakly with environmental density. Massive star-forming galaxies residing in groups are mostly in the red sequence, while lower mass galaxies residing in the field occupy the blue cloud.

These results show that both mass and environment play significant, inter-dependent roles in galaxy evolution, although such roles may become more independent if one neglects potential merger-driven star-forming galaxies. High density regions are linked with triggering high SFRs (for moderate mass galaxies), but this is likely to result in a quicker quenching of star formation as the gas is rapidly consumed, thus quickly transforming groups and cluster environments into the predominant sites for gas-poor, passive galaxies. On the other hand, the quenching processes linked with stellar mass (e.g. AGN) must also play an important role which is distinct from that of environment. It seems that the combination of high masses and high densities is the recipe for very effective and early galaxy formation, as even at z$\sim1$ the star-forming galaxies found with such combination are already on the red sequence, with very low sSFRs (even though most of them are potential mergers).

\end{itemize}

\section*{Acknowledgments}

The authors would like to thank the reviewer for relevant comments and suggestions. DS acknowledges the Funda{\c c}{\~ao para a Ci{\^e}ncia e Tecnologia (FCT) for a doctoral fellowship. PNB acknowledges support from the Leverhulme Trust. IS \& JEG thank the U.K. Science and Technology Facility Council (STFC). The authors acknowledge the fundamental role and unique capabilities of UKIRT and the JAC staff in delivering the extremely high-quality data which allowed this study to be conducted. The authors would also like to thank Tomo Goto, Yusei Koyama, Karina Caputi, Simon Lilly, Jim Dunlop, Carmen Eliche-Moral, Peder Norberg, Len Cowie, Olivier Ilbert, Cheng-Jiun Ma, Ho Seong, Georg Feulner and Francesco Calura for helpful discussions, comments and suggestions. The authors fully acknowledge both the COSMOS and the UKIDSS UDS teams for their tremendous effort towards assembling the extremely high-quality and unique multi-wavelength data-sets.

This article is dedicated to the memory of Timothy Garn (1982-2010).

\bibliographystyle{mn2e.bst}
\bibliography{bibliography.bib}

\appendix

\section[]{The underlying populations} \label{reliability_und}

\subsection[]{Completeness and contamination} \label{comp_und}

State-of-the-art photometric redshifts \citep[][Cirasuolo et al. in prep.]{Ilbert09} can be used to obtain samples of galaxies found at roughly the same redshift as the H$\alpha$ emitters. In this Appendix, the completeness and contamination of the different photometric-redshift-selected samples are investigated and, after correcting for these, the robustness of the results presented in the paper is demonstrated.

For COSMOS, where the photometric redshifts can be obtained more precisely (due to the availability of medium bands which can, for example, probe the 4000 \AA\, break extremely well), samples are defined selecting galaxies with $0.845\pm\Delta z$, with $\Delta z$ ranging from 0.005 to 0.065 ($0.78<z_{\rm photo}<0.91$) in steps of 0.005. Similar samples are defined for the UDS, but in steps of 0.01. The samples are selected over the entire available area with the best photometric and spectroscopic data in both fields.

For the COSMOS field, the $z$COSMOS DR2 secure ($>99$ per cent confidence) spectroscopic redshifts are used to study the completeness (the number of galaxies selected by the photo-$z$ cut that have spectroscopic redshifts within the top hat narrow-band redshift range for the H$\alpha$ emission line [$0.83<z_{\rm spec}<0.86$] compared to the total number of galaxies with spectroscopic redshifts in that range over the same area) and the contamination (the fraction of galaxies in the photo-$z$ sample with spectroscopic redshifts that are confirmed to lie outside the narrow-band redshift range). The photo-$z$ samples are all $\approx10$ per cent spectroscopically complete and there are a total of 260\footnote{From these, 182 are found over the NB$_J$ imaging coverage, with roughly half of these being detected by HiZELS as H$\alpha$ emitters.} spectroscopically confirmed galaxies in the full NB$_J$ filter range (over the entire COSMOS field).

\begin{table}
 \centering
  \caption{A completeness and contamination study of the photometric redshift selected samples obtained in the full COSMOS field using $z$-COSMOS DR2 \citep{zCOSMOS}. Completeness is estimated as the fraction of galaxies wich have spectroscopic redshifts in the narrow-band (top hat) filter range ($0.83<z<0.86$) which are recovered by the photo-$z$ cut (and after accounting for introducing all confirmed galaxies into each sample). The contamination is calculated by computing the fraction of galaxies with spectroscopic redshifts for which the redshifts are not within the narrow-band filter redshift range for H$\alpha$ emitters at $z=0.84$. Photometric-selected samples are all $\approx10$ per cent spectroscopically complete. The samples presented here include H$\alpha$ emitters, but the completeness estimated from excluding the H$\alpha$ emitters is found to be the same. The separating line represents the exact redshift range of the narrow-band filter when assumed to be a perfect top-hat function. Results are also presented in Figure \ref{compl_contamin}. }
  \begin{tabular}{@{}ccccc@{}}
  \hline
   \bf Sample & \bf Sources & \bf Completeness & \bf Contamination \\
   \bf COSMOS & (Number) & C (\%)  & Co (\%)  \\   
  \hline
$0.780<z_{\rm ph}<0.910$ & 14812 & 88 & 73 \\
$0.785<z_{\rm ph}<0.905$ & 14296 & 88 & 72 \\
$0.790<z_{\rm ph}<0.900$ & 13262 & 87 & 70 \\
$0.795<z_{\rm ph}<0.895$ & 11932 & 86 & 68 \\
$0.800<z_{\rm ph}<0.890$ & 11205 & 85 & 66 \\
$0.805<z_{\rm ph}<0.885$ & 10131 & 83 & 65 \\
$0.810<z_{\rm ph}<0.880$ & 9670 & 83 & 64 \\
$0.815<z_{\rm ph}<0.875$ & 8977 & 81 & 63 \\
$0.820<z_{\rm ph}<0.870$ & 8047 & 77 & 59 \\
$0.825<z_{\rm ph}<0.865$ & 6921 & 72 & 53 \\
\hline
$0.830<z_{\rm ph}<0.860$ & 5636 & 65 & 46 \\
$0.835<z_{\rm ph}<0.855$ & 3682 & 49 & 41 \\
$0.840<z_{\rm ph}<0.850$ & 1492 & 41 & 36 \\

  \hline
\end{tabular}
\label{und_testing_table1}
\end{table}

The results for different photo-$z$ cuts can be found in Table \ref{und_testing_table1} and Figure \ref{compl_contamin}. These clearly reveal that adopting a photo-$z$ cut matching the full spectroscopic redshift distribution of the narrow-band filter ($0.825<z_{\rm ph}<0.865$) recovers 72 per cent of the total number of galaxies at those spectroscopic redshifts, missing a significant fraction of the population. On the other hand, while using a wider photo-$z$ cut increases the completeness, it also raises the contamination quite considerably.

\begin{figure}
\centering
\includegraphics[width=7.8cm]{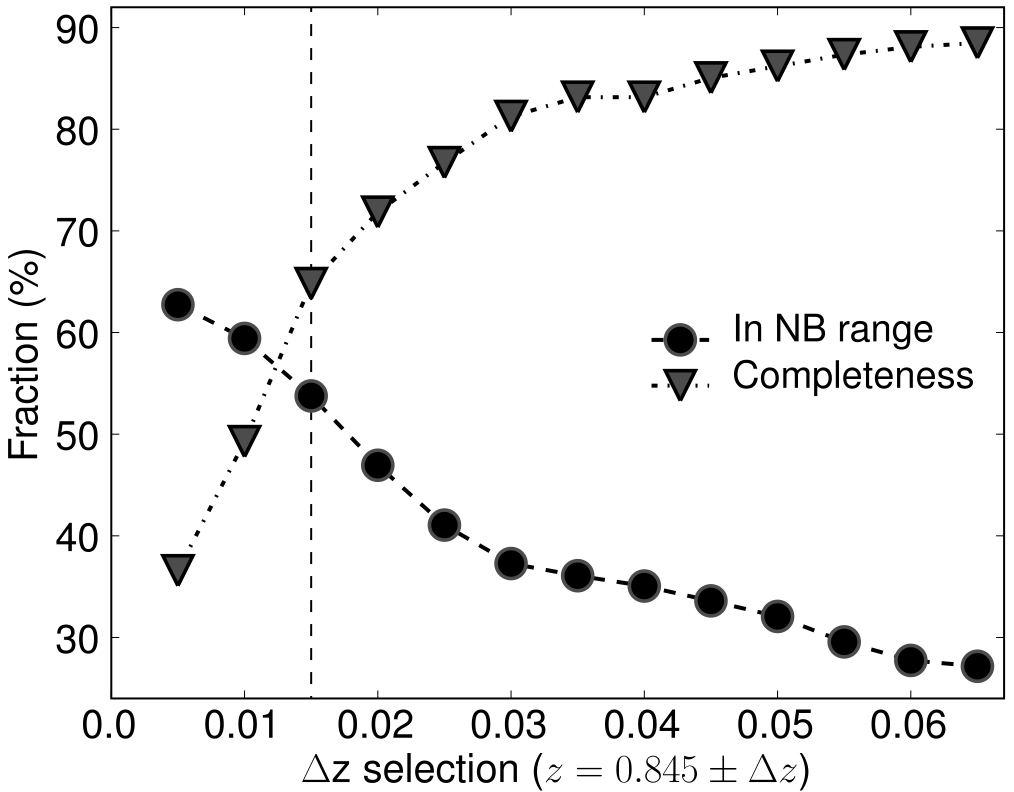}
\caption[wteta]{The completeness fraction and the fraction of sources within the NB filter redshift range ($1-$\,contamination) as a function of photometric redshift width for the COSMOS field. The dashed line indicates the narrow-band filter cut-off assuming a perfect top-hat filter. \label{compl_contamin}}
\end{figure}

It should be noted that in the area covered by the NB$_J$ imaging there are 22 sources spectroscopically confirmed to be within the narrow-band redshift range which are missed by all the studied photo-$z$ cuts (11 of these are H$\alpha$ emitters). These present a median photometric redshift of 0.73 (although some of those have photo-$z$s of $1.3-1.4$), and are much bluer (U-K\,$\approx0.4-3$) than sources recovered by the photo-$z$s (U-K\,$\approx1.5-7.5$); this indicates that the photometric redshift selection is more likely to miss bluer, less massive galaxies than redder, more massive/luminous galaxies. This is mostly a consequence of the fact that blue, star-forming galaxies present weak 4000 \AA \, break features, combined with a significant [OII]3727 emission and Balmer breaks which, combined, can mimic a 4000 \AA \, at lower rest-frame wavelengths and result in lower photometric redshifts. In order to investigate this further, the completeness and contamination are computed independently in different stellar mass bins.
The results, presented in Table \ref{und_testing_table38}, show that the completeness increases as a function of stellar mass; this increase can be well approximated by a 10 per cent increase in completeness per 1 dex in stellar mass for the current sample -- between 9.5 and 11.5 in log M$_{\odot}$. It is also found that the contamination decreases with increasing stellar mass, by approximately 10 per cent per dex in stellar mass. These corrections are incorporated when computing the star-forming fraction as a function of mass.

\begin{table*}
 \centering
  \caption{A completeness and contamination study as a function of stellar mass of the photometric redshift selected samples obtained in the COSMOS field. The results reveal that that while completeness increases as a function of stellar mass, the contamination decreases as a function of mass.}
  \begin{tabular}{@{}ccccccccc@{}}
  \hline
   \bf COSMOS & \bf $10^{9.5}$\,M$_{\odot}$ & \bf $10^{10.0}$\,M$_{\odot}$ & \bf $10^{10.5}$\,M$_{\odot}$  & \bf $10^{11.0}$\,M$_{\odot}$  & \bf $10^{11.5}$\,M$_{\odot}$  \\
   \bf Sample & C (\%) & C (\%) & C (\%) & C (\%) & C (\%)   \\   
  \hline
$0.790<z_{\rm ph}<0.900$ & 56 & 85 & 90 & 97 & 90 \\
$0.800<z_{\rm ph}<0.890$  & 50 & 82 & 89 & 95 & 90 \\
$0.810<z_{\rm ph}<0.880$ & 38 & 81 & 86 & 95 & 90 \\
$0.820<z_{\rm ph}<0.870$ & 25 & 68 & 83 & 93 & 90 \\
  \hline
     \bf Sample &  Co (\%) & Co (\%) & Co (\%) & Co (\%) & Co (\%)   \\ 
       \hline
$0.790<z_{\rm ph}<0.900$ & 80 & 64 & 74 & 68 & 55 \\
$0.800<z_{\rm ph}<0.890$  & 80 & 58 & 71 & 65 & 51 \\
$0.810<z_{\rm ph}<0.880$ & 80 & 52 & 69 & 63 & 51 \\
$0.820<z_{\rm ph}<0.870$ & 80 & 51 & 64 & 55 & 38 \\
  \hline
\end{tabular}
\label{und_testing_table38}
\end{table*}

For the UDS field, the lack of spectroscopic redshifts does not allow a detailed study of the completeness and contamination of the underlying samples; the completeness is therefore estimated using the HiZELS H$\alpha$ sample and the recovery fraction of this sample as a function of the photometric redshift cut. In the COSMOS field, the completeness and contamination estimated using the H$\alpha$ sample are found to be broadly comparable to those from using the population as a whole. Results are presented in Table \ref{und_testing_table2}.

Completeness and contamination may vary as a function of environment as well, particularly because galaxies are spread rather uniformly in the field, but groups are clustered at certain redshifts. Nevertheless, currently the limited spectroscopic samples do not allow to reliably estimate how completeness and contamination may vary with environment, although any such variation is expected to be weak and to largely even out in the statistics, so any possible trends are ignored in this paper.

\subsection[]{The effect of varying photo-z cuts} \label{comp_und}

\begin{figure}
\centering
\includegraphics[width=7.8cm]{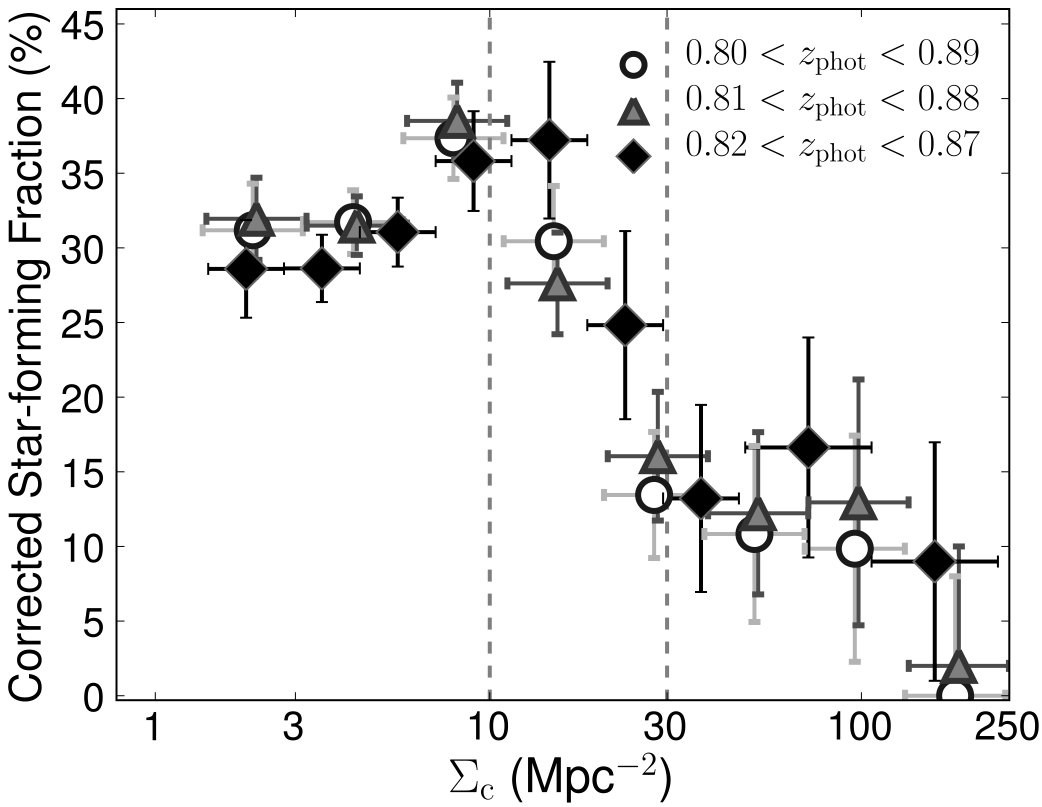}
\caption[wteta]{The relatively small increase of the star-forming fraction within the field regime and fall for denser regions is recovered independently of the photometric redshift cut, once corrections for completeness and contamination are made. Both the underlying population and the estimated local densities are calculated and corrected appropriately for each photo-$z$ sample, but the resultant behaviour of the star-forming fractions as a function of local density for each determination is essentially the same. \label{r10_various_photozsamoples}}
\end{figure}

The choice of the underlying population for estimating local densities can in principle play a significant role in such estimation in a number of ways: sources at significant different redshifts (photo-$z$ outliers) will introduce extra scatter in the measurements; wider photo-$z$ selected samples might pick up large structures just outside the narrow-band filter range and bias certain regions; more importantly, a higher or lower number of sources could change the normalisation of the star-forming fraction. In principal, the latter issue can be avoided by using the completeness and contamination statistics for the underlying samples, calculated in Section A1 to correct the projected local densities ($\Sigma$) and star-forming fractions (SFF) using: \begin{equation}
   \Sigma_{\rm c}=\Sigma\times\frac{1-{\rm Co}}{{\rm C}},\ \ \ \  \rm SFF_c=SFF\times\frac{{\rm C}}{1-{\rm Co}}.
\end{equation} \label{sigma_corr}

\begin{figure}
\centering
\includegraphics[width=7.8cm]{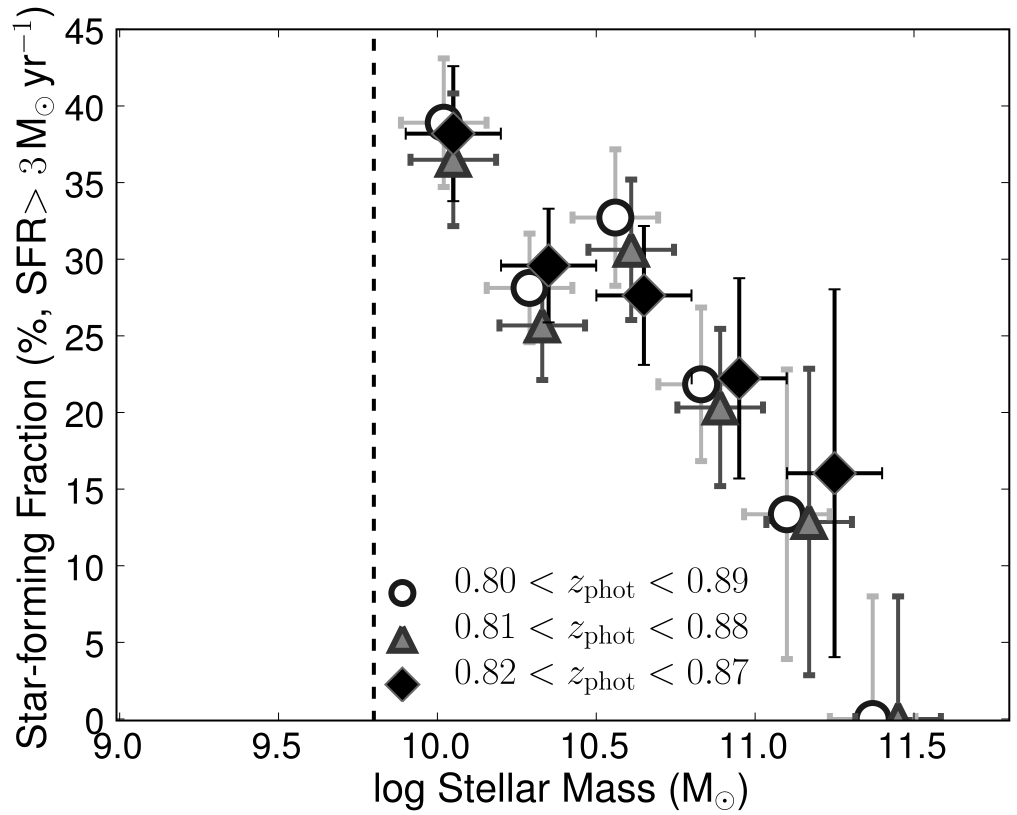}
\caption[wteta]{The mass dependence of the star-forming fraction presented in this paper is clearly recovered independently of the choice of photometric redshift cut. Furthermore, correcting the star-forming fraction based on the mass-dependent completeness and contamination also shows that the normalisation, along with the trend, agrees very well between samples.  \label{mass_dep_diff1}}
\end{figure}

In order to assess any potential biases or errors, corrected local densities have been computed using all of the previously defined underlying populations. It is found that once the corrections are applied, the local density estimates using different underlying samples correlate very well with a slope of unity. Furthermore, as Figure \ref{r10_various_photozsamoples} shows, the trends presented in this paper are independent of the choice of underlying population (within the photo-$z$ range used).

\begin{table}
 \centering
  \caption{A completeness study of the photometric redshift selected samples obtained in the UDS field based on the H$\alpha$ sample. Completeness is estimated as the fraction of H$\alpha$ emitters recovered as compared with the total number. }
  \begin{tabular}{@{}ccccc@{}}
  \hline
   \bf Sample & \bf Sources & \bf Completeness \\
   \bf UDS & (Number) & C (\%)  \\   
  \hline
$0.780<z_{\rm ph}<0.910$ & 3502 & 57 \\
$0.790<z_{\rm ph}<0.900$ & 3078 & 54 \\
$0.800<z_{\rm ph}<0.890$ & 2494 & 49 \\
$0.810<z_{\rm ph}<0.880$ & 1874 & 42 \\
$0.820<z_{\rm ph}<0.870$ & 1502 & 41 \\
$0.830<z_{\rm ph}<0.860$ & 1125 & 34 \\
  \hline
\end{tabular}
\label{und_testing_table2}
\end{table}

Mass-downsizing trends have also been presented in this paper -- these reveal that the fraction of star-forming galaxies decreases as a function of mass. It is fundamental to confirm that this is a consequence of galaxy formation and evolution, rather than a possible bias in either or both the selection of the underlying population and/or the mass estimates. As Figure \ref{mass_dep_diff1} fully details, the mass-downsizing trends are robust against the choice of the underlying population. They are also found to be equally robust against variations in the mass estimates (cf. Appendix B).

\section[]{Stellar Mass Estimates} \label{stellar_mass_rob}

Stellar masses were computed for all H$\alpha$ emitters and galaxies within the underlying population as fully detailed in Section \ref{MK_mass}. Here, further attention is given to the uncertainties and possible systematic errors.

Figure \ref{compa_mobash} shows a comparison between the mass estimates for galaxies in the COSMOS field obtained in this paper and those from \cite{Mobash07}. There is a systematic offset from zero, but this is fully explained by the different Initial Mass Functions (IMF) used: using a Salpeter IMF, as in \cite{Mobash07} results in assuming a mass-to-light ratio higher by a factor of 1.8 when compared to the Chabrier (2003) IMF used here; this results in a difference of $\approx0.17$ dex in stellar mass. The additional scatter is mostly due to the simpler approach used by \cite{Mobash07} to estimate stellar masses from visible luminosities and artificial colours, together with their poorer photometric redshifts. Figure \ref{compa_mobash} also compares the mass estimates for the UDS field with those derived by M. Cirasuolo et al. (in preparation), revealing a lower degree of scatter, which mostly arises from a different aperture correction (M. Cirasuolo et al. use a constant aperture correction assuming all sources are point sources, while in this paper individual corrections are used; the constant aperture correction matches very well to the average of the individual corrections used in this paper), and different stellar synthesis models (with different parameters). Overall, the scatter between the different determinations is relatively small and there are no systematic trends.

\begin{figure}
\centering
\includegraphics[width=7.2cm]{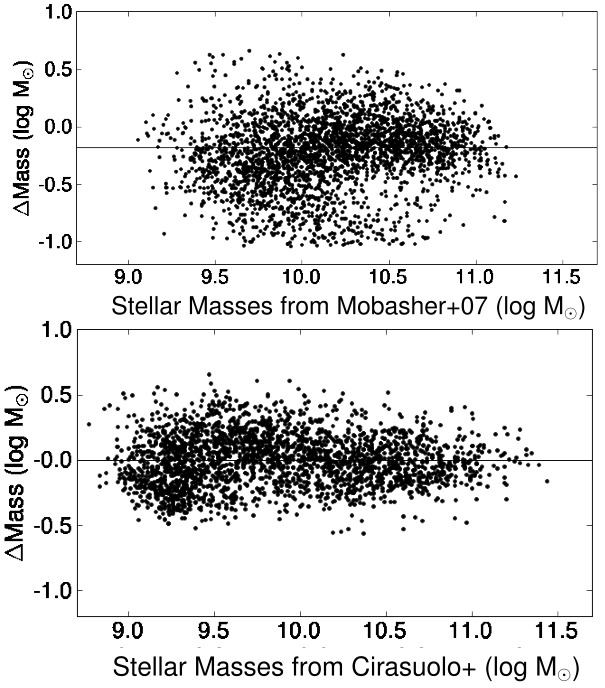}
\caption[wteta]{Comparison between the determined stellar masses and those from \cite{Mobash07} and Cirasuolo et al. (in prep.). The ones computed in this study, by using a much larger number of bands and by taking a full SED fitting approach fixing the redshift are much more reliable than those estimated from rest-frame $V$ luminosities combined with artificial rest-frame B-V colours by \cite{Mobash07} (those authors also use a Salpeter IMF; this results in estimating stellar masses which are $+0.17$ dex higher). Nevertheless, there is no significant offset between both estimates (once one corrects for the different IMFs), and the scatter is fully consistent with the errors expected from the Mobasher et al determinations. As for the comparison with the preliminary results from Cirasuolo et al., the differences (scatter) mostly arise from the use of different models, different aperture corrections and, to a lower extend, the use of $z=0.845$ in this paper instead of the best photo-z.  \label{compa_mobash}}
\end{figure}

\bsp

\label{lastpage}

\end{document}